\documentclass[%
 reprint,
groupedaddress,
 amsmath,amssymb,
 aps,
]{revtex4-1}

\usepackage{graphicx}
\usepackage{dcolumn}
\usepackage{bm}
\usepackage[
            pdfstartview=FitH,
            CJKbookmarks=true,
            bookmarksnumbered=true,
            bookmarksopen=true,
            colorlinks, 
            linkcolor=blue,
            anchorcolor=blue,
            citecolor=blue,
            ]{hyperref}

\makeatletter
\renewcommand{\section}{\@startsection{section}{1}{0mm}{-\baselineskip}{0.4\baselineskip}{\bf\leftline}}
\makeatother

\makeatletter
\renewcommand\frontmatter@abstractwidth{\dimexpr\textwidth-0.5in\relax}
\makeatother

\begin{document}

\preprint{APS/123-QED}

\title{Ultrahigh charge electron beams from laser-irradiated solid surface}

\author{Yong Ma$^{1, 7}$}
\author{Jiarui Zhao$^{1}$} 
\author{Yifei Li$^{1}$}
\author{Dazhang Li$^{4}$}
\author{Liming Chen$^{1, 2, 3}$}
\email{lmchen@iphy.ac.cn}
\author{Jianxun Liu$^{5}$}
\author{Stephen J. D. Dann$^{7}$}
\author{Yanyun Ma$^{5}$}
\email{yanyunma@126.com}
\author{Xiaohu Yang$^{5}$}
\author{Zheyi Ge$^{5}$}
\author{Zhengming Sheng$^{2, 6}$}
\author{Jie Zhang$^{2, 3}$}
\email{jzhang1@sjtu.edu.cn}

\affiliation{$^{\textbf{1}}$Institute of Physics, Chinese Academy of Sciences, Beijing 100190, China}
\affiliation{$^{\textbf{2}}$IFSA Collaborative Innovation Center and Department of Physics and Astronomy, Shanghai Jiao Tong University, Shanghai 200240, China}
\affiliation{$^{\textbf{3}}$School of Physical Sciences, University of Chinese Academy of Sciences, Beijing 100049, China}
\affiliation{$^{\textbf{4}}$Institute of High Energy Physics, Chinese Academy of Sciences, Beijing 100049, China}
\affiliation{$^{\textbf{5}}$College of Science, National University of Defense Technology, Changsha 410073, China}
\affiliation{$^{\textbf{6}}$Department of Physics, Scottish Universities Physics Alliance, University of Strathclyde, Glasgow G4 0NG, United Kingdom}
\affiliation{$^{\textbf{7}}$Department of Physics, Lancaster University, Bailrigg, LA1 4YW, United Kingdom}




\begin{abstract}

Compact acceleration of a tightly collimated relativistic electron beam with high charge from a laser-plasma interaction has many unique applications. However, currently the well-known schemes, including laser wakefield acceleration from gases and vacuum laser acceleration from solids, often produce electron beams either with low charge or with large divergence angles.  In this work, we report the generation of highly collimated electron beams with a divergence angle of a few degrees, quasi-monoenergetic spectra peaked at the MeV level, and extremely high charge ($\sim$100 nC) via a powerful sub-ps laser pulse interacting with a solid target in grazing incidence. Particle-in-cell simulations illustrate a new direct laser acceleration scenario, in which the self-filamentation is triggered in a large-scale near-critical-density plasma and electron bunches are accelerated periodically and collimated by the ultra-intense electromagnetic field. The energy density of such electron beams in high-Z materials reaches to $\sim10^{12} \mathrm{J/m^{3}}$, making it a promising tool to drive warm or even hot dense matter states.

\end{abstract}

\maketitle


\section*{INTRODUCTION}

In studies of laser-plasma acceleration (LPA), several laser wakefield accelerator (LWFA)  \cite{tajima1979} concepts have been proposed in the last few decades, including the plasma beat wave accelerator \cite{tajima1979, rosenbluth1972}, self-modulated laser wakefield accelerator (SM-LWFA)  \cite{krall1993}, cross-modulated laser wakefield accelerator (XM-LWFA)  \cite{sheng2002} and LWFA in the bubble regime \cite{pukhov2002, kostyukov2004}. 
The successful generation of high quality electron beams at GeV scale with quasi-monoenergetic spectra stimulates the study of laser-plasma accelerators worldwide \cite{mangles2004, geddes2004, faure2004, leemans2006, hafz2008, wang2013, kim2013, leemans2014}. 
However, almost all LPA experiments and theoretical models are based on interactions between lasers and gases, limiting the beam charge to typically a few tens of pC.
While the charge of the electron bunch could reach a few nC in laser-solid interactions due to higher absorption efficiency and attempts have been made to optimize beam collimation \cite{chenlm2001, liyt2001, liyt2006, habara2006, mordo2009, tian2012, mao2012, Wang2013578}, the beam quality still needs to be greatly improved due to large divergence angles and quasi-thermal broad energy spectra.
Such electrons are usually generated via several heating mechanisms such as resonant absorption \cite{estabrook1978}, vacuum heating \cite{brunel1987,  chenlm2001-2}, J$\times$B heating \cite{kruer1985}, and stochastic heating \cite{sheng2002-2}. 
Directional electron beams with nC charge have been produced via vacuum laser acceleration (VLA) with a plasma mirror injector \cite{thevenet2016}. 
Unfortunately, the beam collimation also suffers from the pondermotive force of the laser pulse in vacuum during acceleration, which results in a large divergence angle (hundreds of mrads) and a halo in the electron beam profile. 
Recently, few MeV quasi-monoenergetic electron acceleration has been observed in fs laser-solid interaction with beam divergence angles of $\mathrm{1-2^{o}}$ \cite{mao2015}. 
However, the beam charge is still limited to hundreds of pC, and the underlying physics of such acceleration remains unclear. 

In this work, electron beams with extremely high beam charge of approximately 100 nC are generated for the first time in 200 TW, sub-ps laser-solid interactions with deliberately induced pre-plasma. 
The electron beams are highly collimated with an average divergence angle $<\mathrm{3^{o}}$ and the energy spectra are quasi-monoenergetic with peaks at several MeV.

PIC simulations illustrate a new scenario of electron acceleration in which the acceleration and confinement regimes are combined in a unique way. 
It is shown that electron beams are mainly produced via direct laser acceleration (DLA) \cite{pukhov1999, gahn1999, mangles2005, kneip2008, liyy2011, zhang2015, huang2016} in plasma channels~\cite{louise2011, louise2013} driven by the long laser pulse in a large scale near-critical pre-plasma. 
The strong electromagnetic field inside the plasma channel confines the electron beams tightly,  preserving the collimation of the beam. 
The significant improvement of the beam charge benefits from the direct energy transition from laser pulse into electron beams during the persistently DLA process. 

\section*{Experimental Results}

\begin{figure}[b]
\centering
\includegraphics[width=.8\linewidth]{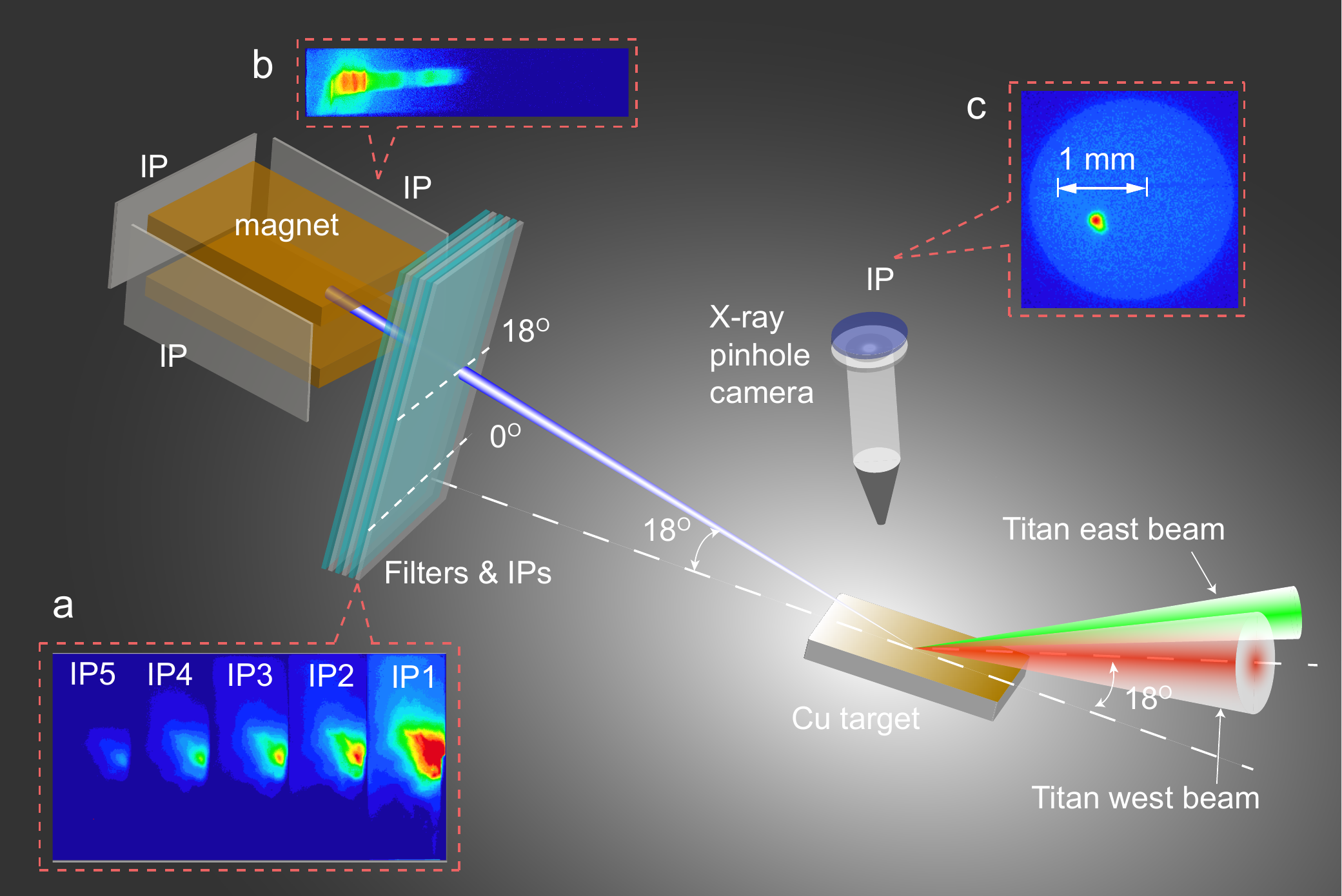}
\caption{Experimental set up. Inset figures present (a) the angular distribution of the electron beam on image plates; (b) the energy distribution after deflected by the spectrometer; (c) the image of the X-ray source detected by an X-ray pinhole camera. }
\label{fig:fig1}
\end{figure}

The experiment was performed on Titan at the Jupiter Laser Facility at Lawrence Livermore National Laboratory (LLNL). 
The setup of the experiment is shown in Fig.~\ref{fig:fig1} (see \textbf{METHOD} for details). 
Copper block targets were irradiated by a 200 TW, sub-ps laser at an incident angle of $\mathrm{72^{o}}$ in P-polarization. 
The laser pedestal 3 ns prior to the main pulse (pre-pulse) was measured to be 5$\pm$2 mJ at 1$\omega$ with full laser energy of 150 J and $\sim0.2~\mu$J at 2$\omega$ with full laser energy of 30$\pm$5 J. 

\begin{figure}[b]
\centering
\includegraphics[width=.8\linewidth]{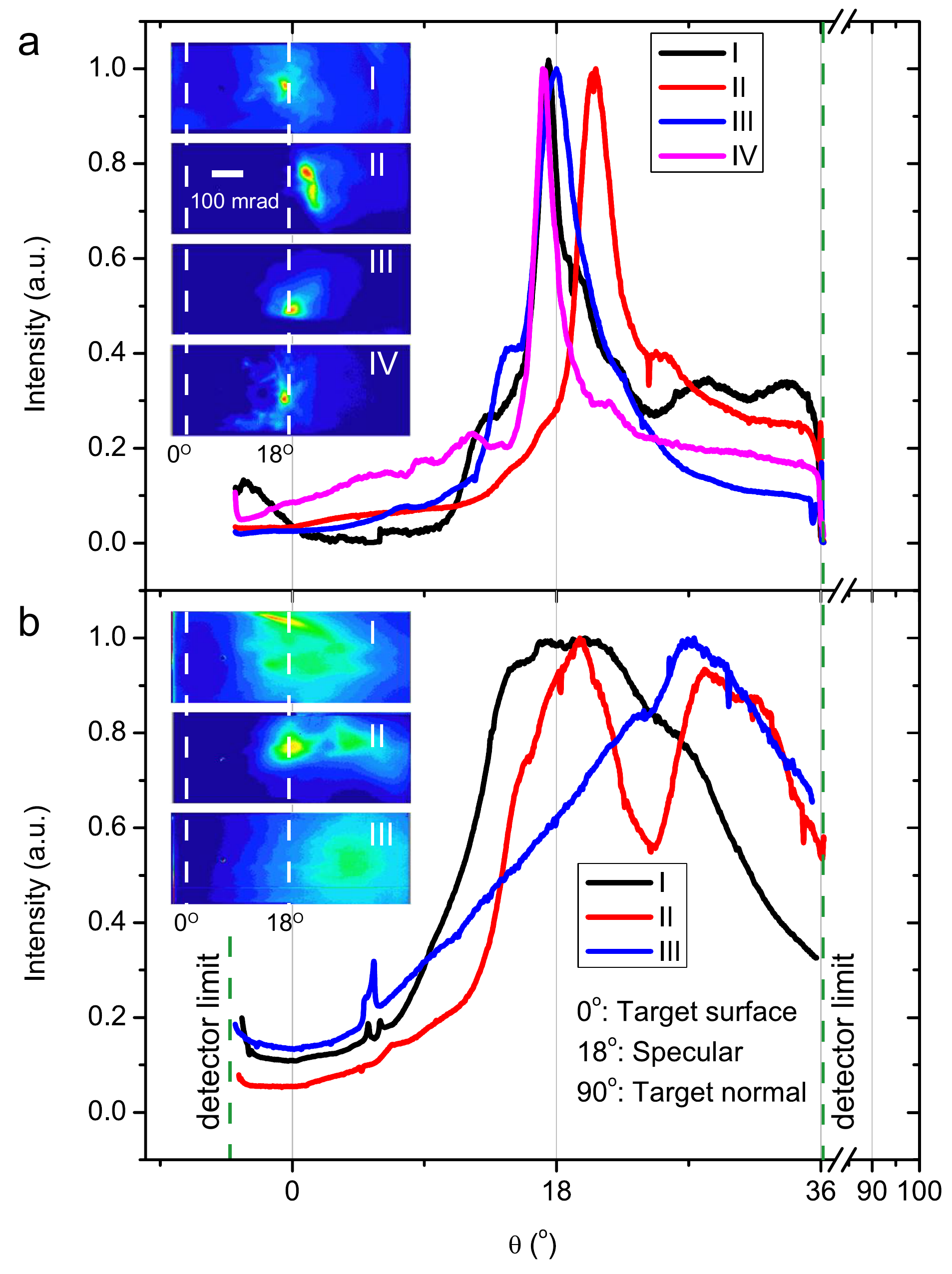}
\caption{Angular distribution of the electron beams with 1$\omega$ 150 J laser pulse. Pre-pulse energy in a(I-IV) are 7, 5, 7, 4 mJ, respectively; Pre-pulse energy in b(I-III) are 20, 32, 82 mJ, respectively.  }
\label{fig:fig2}
\end{figure}

\begin{figure}
\centering
\includegraphics[width=.8\linewidth]{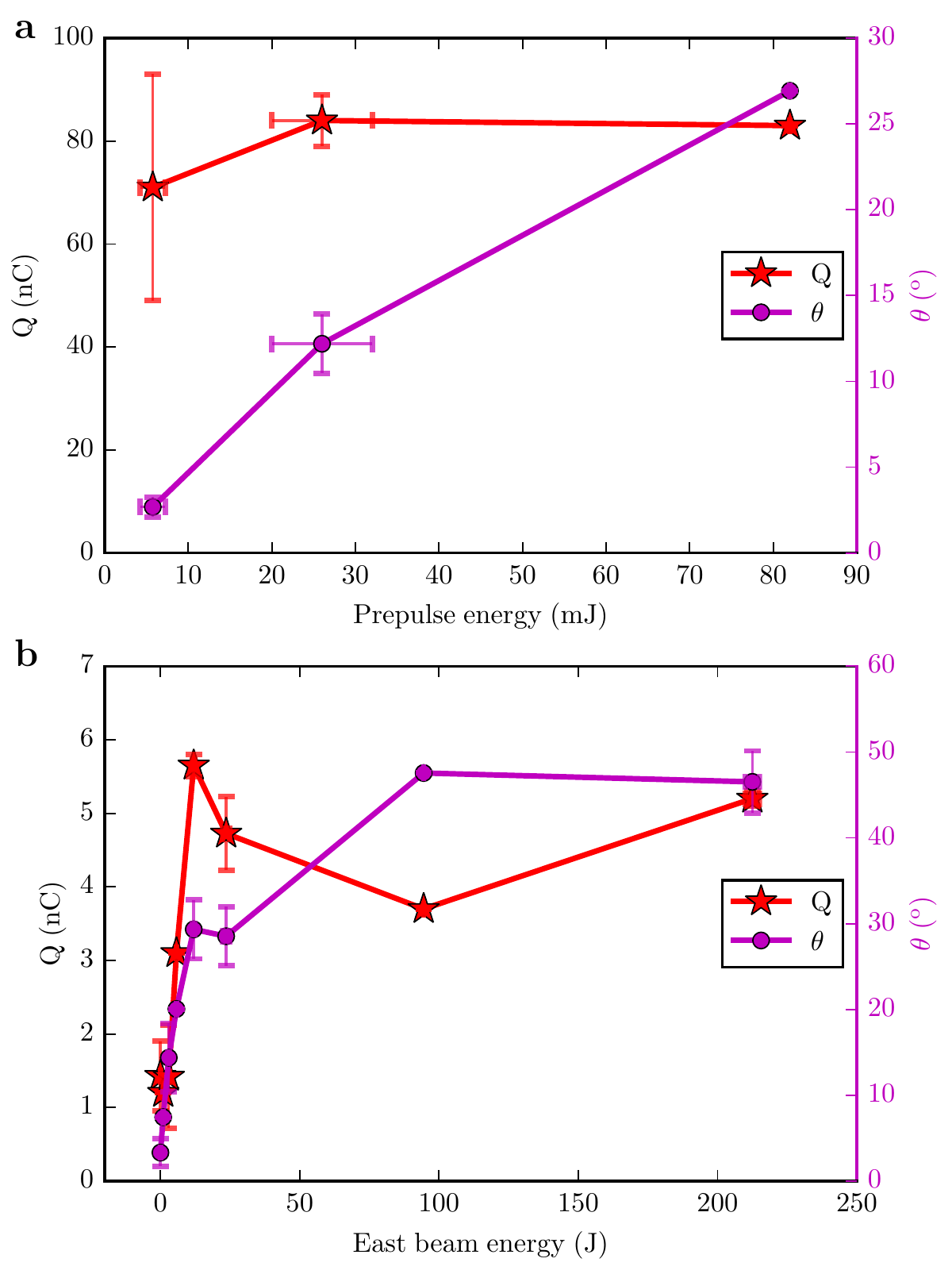}
\caption{Electron beam charge and divergence angle. (a) Dependence of electron beam charge and divergence angle on the intrinsic pre-pulse energy of the 1$\omega$ main pulse at 150 J; (b) Dependence of electron beam charge and divergence angle on pre-pulse (Titan east beam) energy with 2$\omega$ main pulse at 30$\pm$5 J.  }
\label{fig:fig3}
\end{figure}

\begin{figure}
\centering
\includegraphics[width=.8\linewidth]{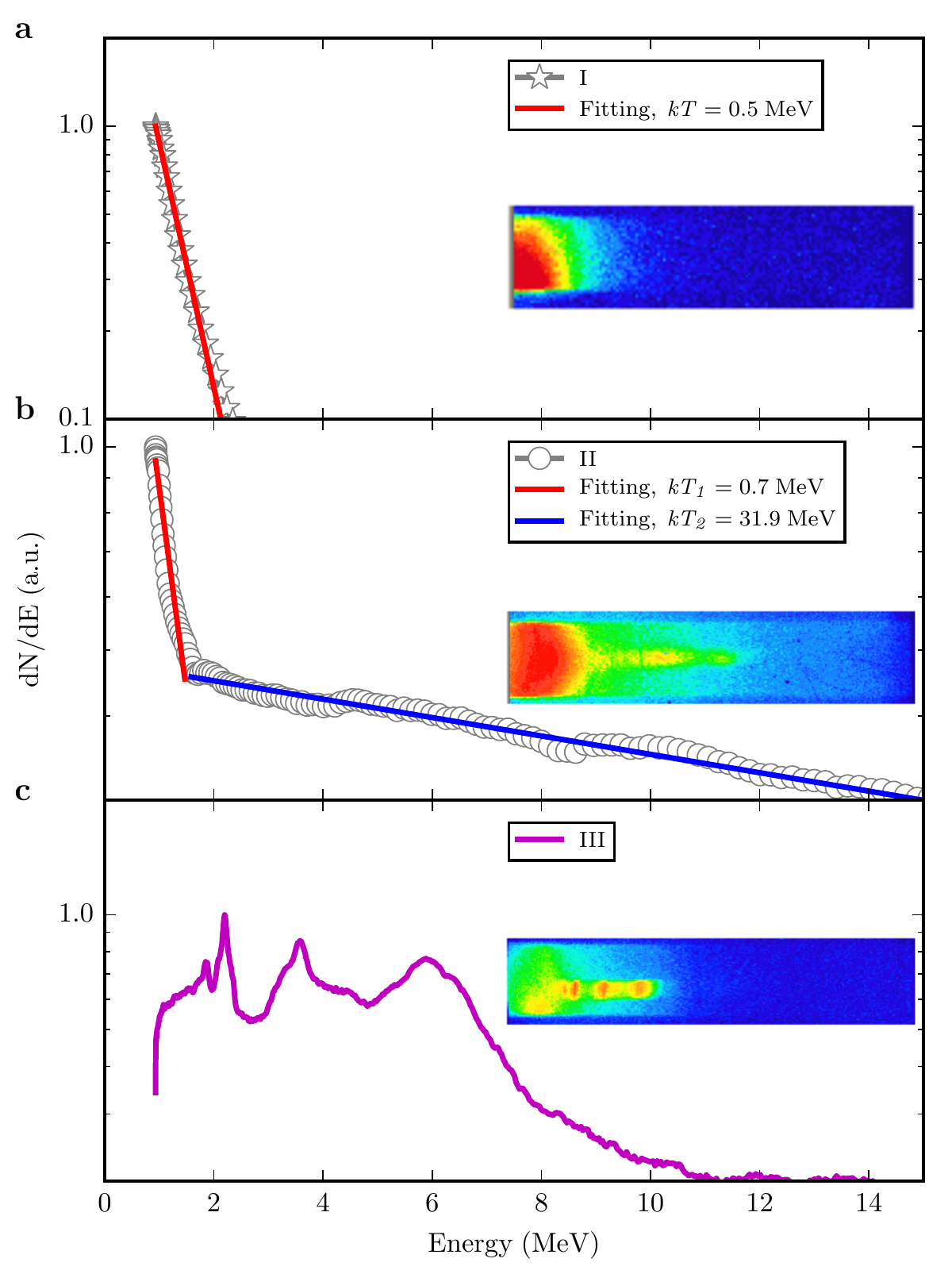}
\caption{Energy spectra of the electron beams with different laser parameters. (a) High contrast 2$\omega$ under the main pulse energy of 30$\pm$5 J; (b) 1$\omega$ with high intrinsic pre-pulse energy under the main pulse energy of 150 J; (c) 1$\omega$ with low intrinsic pre-pulse energy under the main pulse energy of 150 J. }
\label{fig:fig4}
\end{figure}

Highly collimated electron bunches with good pointing stability and extremely high beam charge were generated, as shown in Fig.~\ref{fig:fig2}a, using the full energy laser pulse with approximately 5 mJ pre-pulse.
These beams were emitted along the laser specular reflection direction with an average divergence angle of  $\mathrm{2.7^{o}}$ FWHM.
This is much smaller than those generated via the VLA mechanism in laser-solid interactions but similar to the cases of laser- driven wakefield acceleration from gas.
The beam charge can be as high as 94 nC with energy above 1.0 MeV (see Fig.~\ref{fig:fig2}a-III and Fig.~\ref{fig:fig3}a). 
To the best of our knowledge, this is the first observation of electron bunches with such high charge and such a small divergence angle. 
The beam current reaches $I \simeq 134$ kA by assuming the pulse length of the electron beam is the same as that of the laser pulse.
This is a large fraction of the Alf\'ven current limit \cite{alfven1939, dodin2006}, which in this case is $I_{A} = 1.65\times 17 \mathrm{[kA]}\sqrt{\gamma^{2}-1} = 262$ kA, where $\gamma = 9.4$ for the average energy of 5.3 MeV, as shown in Fig.~\ref{fig:fig4}c. 

In addition to the generation of collimated beams (the central bright spot) in the laser specular direction, a weak plateau appears between laser specular and target normal. 
This indicates that the generation mechanism of the plateau electrons differs from that of the central bright spot and the energy of such electrons could be much lower. 
The outgoing direction of plateau electrons is energy dependent: $\sin\alpha'=(\gamma-1/\gamma)\sin\alpha$~\cite{sentoku1999}, where $\gamma$ is the Lorenz factor of electrons, $\alpha'$ and $\alpha$ are angles between the target normal and the outgoing and laser specular directions, respectively. 

When increasing the pre-pulse energy to a few tens of mJ, the electron beam divergence increases significantly but with a similar level of beam charge, as shown in Fig.~\ref{fig:fig2}b and Fig.~\ref{fig:fig3}a. 
In this case, the outgoing direction of most electrons is between the laser specular and target normal directions. 
Note that the peak intensity of the 5 mJ pre-pulse reaches $9\times 10^{15}\mathrm{W/cm^{2}}$, which is already high enough to produce pre-plasma on the solid target. 
It seems that the pre-plasma scale length has an adverse effect on the beam collimation, but not on the beam charge. 

To further investigate the influence of the pre-plasma, we also gradually reduced the pre-pulse energy, and hence the pre-pulse intensity.
However, the 5 mJ pre-pulse is the smallest which can be achieved with this laser operating at its fundamental frequency (1$\omega$). 
Therefore, second harmonic (2$\omega$) laser pulses were used, lowering the pre-pulse energy to  $\sim0.2~\mu$J with intensity below the ionization threshold.
A ns laser pulse (Titan east beam in Fig.~\ref{fig:fig1}) was used as an additional pre-pulse to produce pre-plasma. 
The dependence of the electron beam divergence angle and the beam charge on the additional pre-pulse energy are shown in Fig.~\ref{fig:fig3}b. 
Before the introduction of the pre-pulse, i.e., using the 2$\omega$ Titan west beam only, the outgoing direction of the electron beams is still along laser specular and the average divergence angle is $\mathrm{3.3^{o}}$, which is very close to that of Fig.~\ref{fig:fig2}a.
However, the beam charge decreases to an average of $\sim$1.5 nC because of much lower laser energy. 
Then, increasing the energy of the pre-pulse gradually and keeping the main pulse energy fixed at $\sim$30 J, the average FWHM divergence angle of the electron beams increases accordingly and reaches to a maximum of  46.4$\mathrm{^{o}}$, while the beam charge quickly increases and then remains at the same level ($\sim$5 nC on average).
This dependence of beam divergence on pre-pulse energy with 30 J 2$\omega$ drive laser is similar to that with 150 J 1$\omega$ drive laser. 
We conclude that even though the main pulse energy plays the key role in controlling the electron beam charge, the pre-plasma condition has significant effects on the divergence angle of the electron beam.

The energy spectra of the out-going electron beams are shown in Fig.~\ref{fig:fig4}. 
In Fig.~\ref{fig:fig4}a, with the 2$\omega$ laser pulse and no pre-plasma, most of the electrons are low energy ($<$1 MeV) and the exponential decay fitting gives an effective temperature $kT$ = 0.5 MeV; 
In the case of the 1$\omega$ laser pulse with high pre-pulse, the energy spectrum in Fig.~\ref{fig:fig4}b demonstrates an obvious dual-temperature distribution. 
Although the majority are still below 1 MeV with an effective temperature of $kT_{1}$ = 0.7 MeV, the high-energy tail reaches 20 MeV with a much higher effective temperature of $kT_{2}$ = 31.9 MeV. 
It is obvious that these two groups of electrons with different temperatures have been generated via different mechanisms. 
Low temperature electrons might be produced by a laser heating process, such as resonant absorption, J$\times$B heating and so on. 
The generation of high temperature electrons could be a result of a particular acceleration process (rather than heating). 
When lowering the pre-pulse energy of the 1$\omega$ laser pulse to 5 mJ, the spectrum becomes quasi-monoenergetic with peaks at 2-6 MeV and the amount of lower energy electrons is greatly suppressed, as shown in Fig.~\ref{fig:fig4}c.
These are the same laser parameters as in Fig.~\ref{fig:fig2}a where tightly collimated electron beams with extremely high beam charge were observed.

\section*{Simulation and discussion}

To investigate the mechanism of the generation of such collimated electron beams with quasi-monoenergetic spectra and extremely high beam charge, PIC simulations (see \textbf{METHOD}) have been performed and the results agree qualitatively with those of the experiment. 

\begin{figure*}[t]
\centering
\includegraphics[width=.98\linewidth]{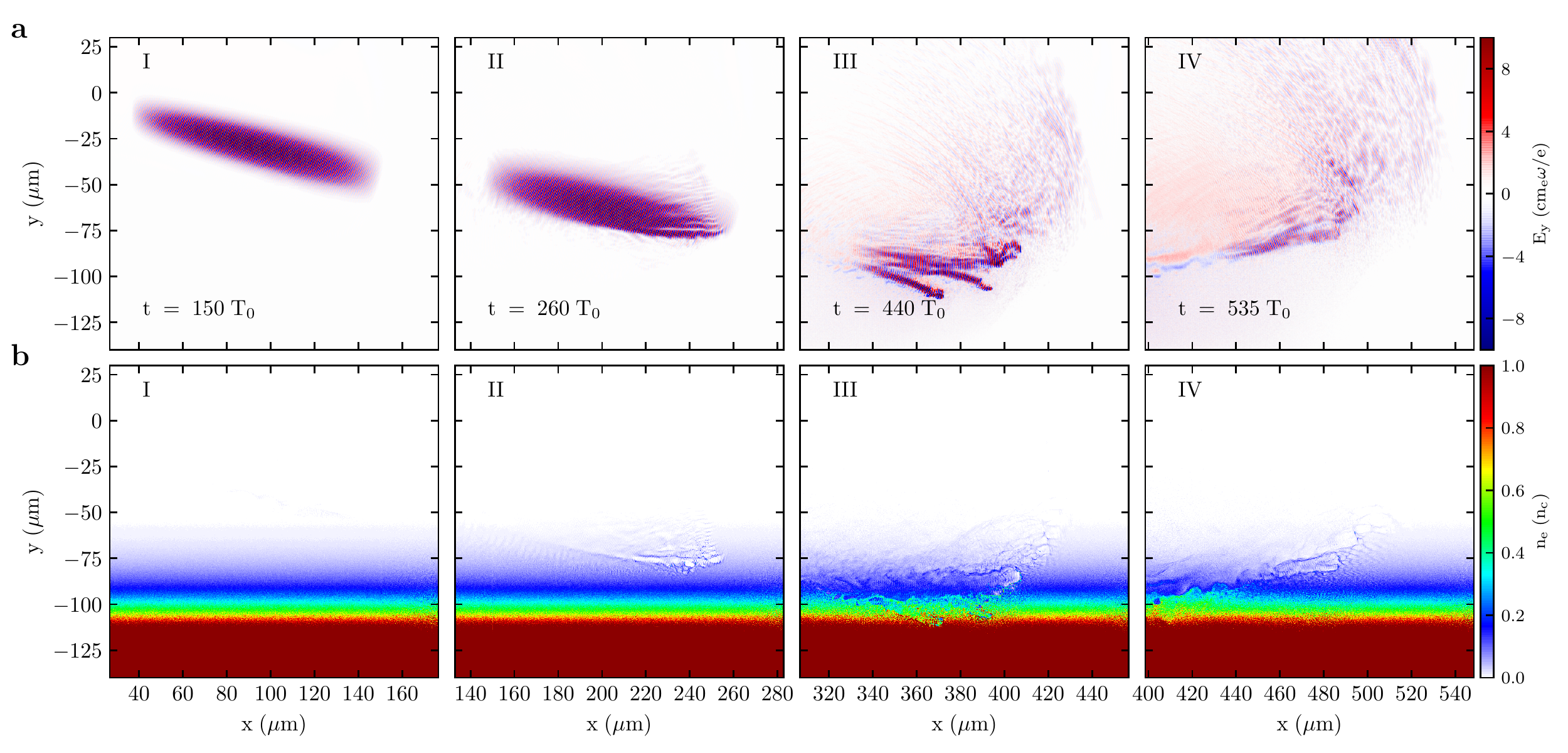}
\caption{Snapshots of laser fields and plasma electron density distributions at 4 time steps obtained from PIC simulation. (a) Laser field distributions; (b) Plasma electron density distributions..}
\label{fig:fig5}
\end{figure*}

The general scenario of the interaction is shown in Fig.~\ref{fig:fig5}, where Fig.~\ref{fig:fig5}a gives the evolution of laser field in the plasma and Fig.~\ref{fig:fig5}b gives the evolution of plasma density. 
The self-filamentation process is enhanced by grazing incidence. 
As the laser pulse penetrates into the near-critical-density region, the lower part of the beam, which is in the higher plasma density, is reflected by the plasma and interacts with the less affected upper part in the relatively lower density. 
As a consequence, the superposition of these two parts leads to a transverse self-modulation in intensity, i.e. self-filamentation, as shown in Fig.~\ref{fig:fig5}a(II).  
As the laser pulse penetrates further into the higher density region, the laser pulse breaks up into 3 main filaments. 
As shown in Fig.~\ref{fig:fig5}a(III) at $t = 440~T_{0}$, the top filament starts to be reflected and the other two keep penetrating into the overdense plasma.
All 3 filaments drive their own plasma channels, as shown in Fig.~\ref{fig:fig5}b(III). 
However, the two lower ones disappear eventually after the energy is fully depleted. 
The upper filament survives and propagates along the laser specular direction where it continuously drives its plasma channel, trapping and heating electron bunches as shown in Fig.~\ref{fig:fig5}a(IV) and Fig.~\ref{fig:fig5}b(IV). 
The electron bunching with constant spacing in the plasma channel indicates that the acceleration mechanism is similar to DLA. 

\begin{figure}[t]
\centering
\includegraphics[width=.98\linewidth]{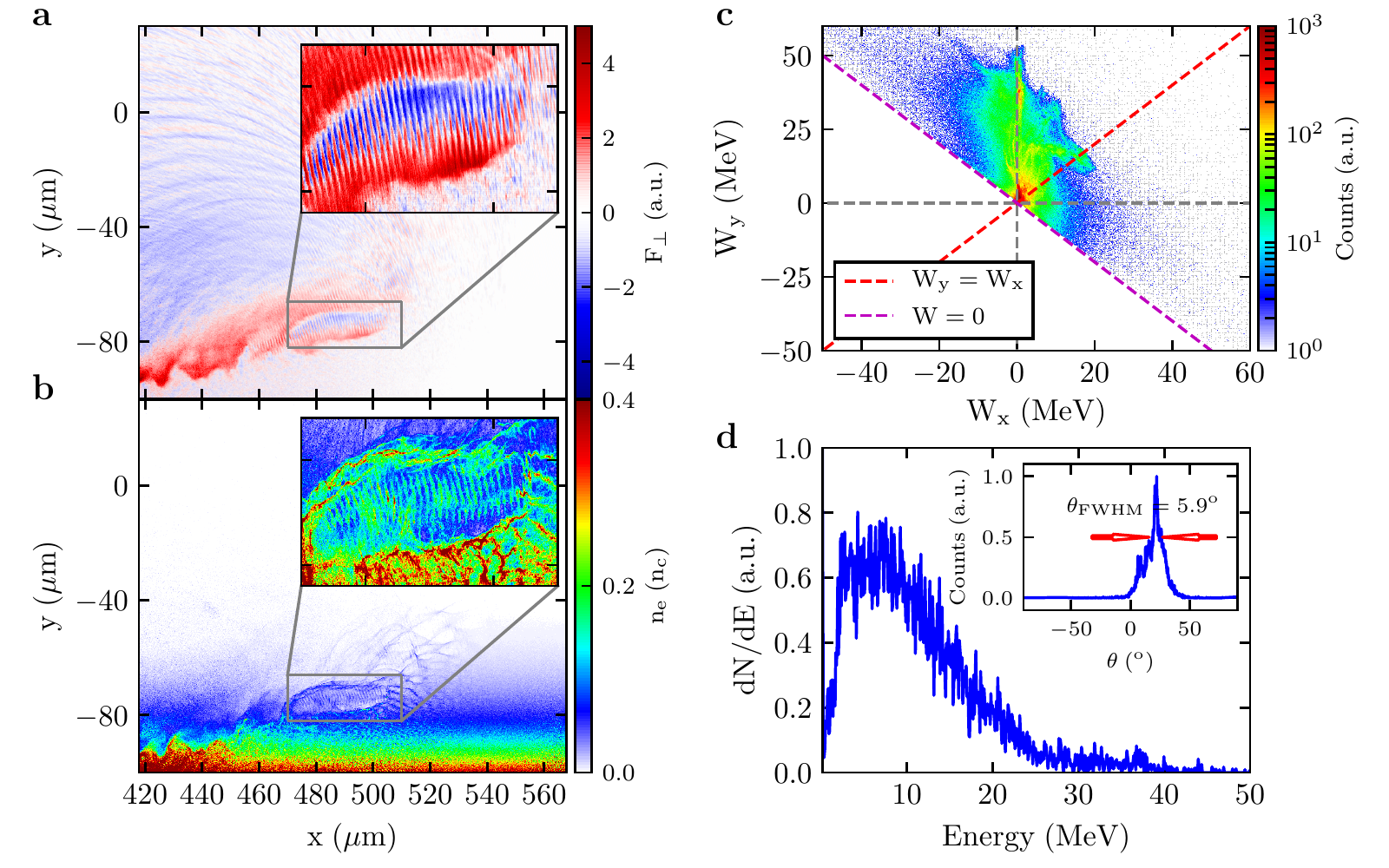}
\caption{(a), (b) show the transverse force and plasma density distribution at $t = 555~T_{0}$, respectively. The inset figures in (a) and (b) show the transverse focusing force and the fine structure of the electron beam distribution inside a plasma channel. The color scale of the inset figures are reduced by a factor of 10. (c) The energy gain components distribution in ($W_{x}-W_{y}$) space at $t = 555~T_{0}$. The red dashed line divides the space into two regions: DLA dominated region in the upper left and the wakefield acceleration dominated region in the lower right. Electrons above the horizontal grey dashed line gain energy in the laser field while those below lose energy. Electrons to the right of the vertical dashed line gain energy from the wakefield while those to the left lose energy. . (d) The energy spectrum of electrons escaping from the plasma at a slightly later time at $t = 585~T_{0}$. The inset shows the corresponding angular distribution of the electron beam. }
\label{fig:fig6}
\end{figure}

To deeply understand the strong collimation of the electron beam, the transverse electromagnetic force $F_{\perp} \sim E_{y} - cB_{z}$ is given in Fig.~\ref{fig:fig6}a. 
The inset of Fig.~\ref{fig:fig6}b illustrates that the overall electromagnetic force inside the plasma channel (see inset of Fig.~\ref{fig:fig6}b) tends to focus the electron beam, which results in the self-collimation of the electron beam.
Similar phenomena were also found in Ref.~\cite{liub2013}.

To understand the detailed procedures of the acceleration, the electron distribution in energy gain space of ($W_{x}-W_{y}$) at $t=555~T_{0}$ are given in Fig.~\ref{fig:fig6}c. 
Here $W_{x}$ and $W_{y}$ are the work done on the electrons by the $x$ and $y$ components of the electric field. 
Moreover, $W_{y}$ and $W_{x}$ are respectively, the energy gain from the laser field which represents the DLA, and the energy gain from the electrostatic field along the laser propagating direction which represents wakefield acceleration. 
It is very clear that the dominant acceleration mechanism is DLA since most of the electrons are located in the region where  $W_{y} > W_{x}$. 

As a consequence of the collimation and acceleration inside the plasma channel, the electron spectrum agrees with that of the experiment, as shown in Fig.~\ref{fig:fig6}d.
The simulated electron beam propagates along $\mathrm{22.1^{o}}$ from the $x$ axis, very close to the laser specular direction, with a FWHM divergence angle of $\mathrm{5.9^{o}}$ which is similar to experiment, as shown in the inset of Fig.~\ref{fig:fig6}d.


Our experimental observation can expel another electron acceleration mechanism, VLA, in laser-solid interaction.
In the mechanism of VLA, during the direct interaction with the laser field, electrons will escape the focal volume transversely after gaining enough transverse momentum, resulting in large divergence angle of the beam. 
Additionally, the transverse pondermotive force also tends to expel electrons from the laser axis and leads to the formation of a hollow structure in the profile of the electron beams, as observed in Ref.~\cite{mordo2009, tian2012,thevenet2016}. 
However, the electron beams in our experiment are tightly collimated with small divergence angle $\sim\mathrm{2.7^{o}}$ and without the hollow structure. 
This reveals the importance of the self-filamentation process and the corresponding channelling process in preserving the collimation of the high charge electron beam. 

DLA in a high density plasma channel from solid is also different from LWFA in low density plasmas, especially the so-called bubble regime in which the acceleration mainly occurs in the first wave bucket. 
In LWFA, the beam charge is limited to a few hundred pC due to the beam-loading effects which follow $Q\propto(k_{p}R_{b})^{4}/\sqrt{n_{e}}$~\cite{tzoufras2008} , where $R_{b}$ is the bubble radius. 
We believe this is the main reason for the low beam charge in Ref.~\cite{mao2015} when fs laser pulse incident on large scale but lower density plasma from solid surface and form a single bubble. 
In DLA driven by ps laser pulse, without the limitation of beam-loading, a separate bunch of electrons can be driven in each half optical cycle. 
The total beam charge in simulation is proportional to the number of electron bunches in the plasma channel. 
The long laser pulse duration provides the energy required to sustain the continuous acceleration, and this is in accordance with the fact that the beam charge increases as the laser energy increases in experiment. 

Such high charge and high current beams may be used to drive high energy states of matter. 
Considering the penetrating depth $\sim$1 mm in Au, for example, the energy density of the electron beams reaches $4\times10^{12} \mathrm{J/m^{3}}$, even higher than that of the LCLS XFEL beam which has been proven as a powerful tool to drive warm dense matter states~\cite{LCLS}. 
Note that the attenuation length of MeV electrons is much longer than that of optical laser and XFEL, which makes it an ideal tool to drive warm dense matter with large scale.

\section*{Conclusions}
In conclusion, by using 200 TW sub-picosecond laser pulses, tightly collimated ($\sim\mathrm{2.7^{o}}$), directional and quasi-monoenergetic MeV electron beams with extremely high beam charge ($\sim$100 nC scale) were generated experimentally for the first time. 
We found that the generation of such electron beams relies on the laser contrast and laser energy.
PIC simulations illustrate a new electron acceleration scenario in laser-solid interaction. 
As the laser pulse propagates in the near-critical-density plasma, the self-filamentation process drives the formation of a bubble-like plasma channel, which confines the laser filament itself.
Electrons are accelerated via DLA in each optical cycle and confined in a small region inside the plasma channel due to the ultra-intense electromagnetic focusing force.
In the case of long pulse duration with many optical cycles, the energy transfer from laser pulse to electron beams boosts the beam charge significantly. 
Such a high charge electron accelerator might find wide applications in seeding high flux ($\sim2\times10^{11}$photons/ps) $\gamma$-ray sources, single shot electron radiography and even in the fast ignition concept \cite{tabak1994}. Most importantly, the extremely high energy density of such electron beam makes it a promising pump for warm/hot dense matters.\\

\setlength{\parindent}{0pt}
\textbf{METHOD}

\setlength{\parindent}{0pt}
\textbf{Laser system.} The experiment was performed on Titan at the Jupiter Laser Facility at Lawrence Livermore National Laboratory (LLNL). 
Titan is a two-arm laser system with a sub-ps west beam and a ns east beam. 
The wavelength of both arms is 1053 nm. 
The west beam was used as the main pulse, with total energy of 150 J in 700 fs full-width at half maximum (FWHM) pulse duration. 
It was focused by an f/3.5 off-axis parabola to a 7 $\mu$m $1/e^{2}$ spot size, resulting in a laser intensity of  $2.8 \times 10^{20}\mathrm{W/cm^{2}}$ ($a_{0}$=15). 
The laser pedestal measured at 3 ns prior to the main pulse, i.e. the intrinsic pre-pulse, was $5\pm2$ mJ.

\setlength{\parindent}{12pt}
By using a KDP crystal for second harmonic of laser pulse, the pre-pulse energy can be decreased to 0.2 $\mu$J, while the energy of the main pulse is reduced to 30$\pm$5 J. 
The east beam was used as an additional pre-pulse when the main pulse was at 2$\omega$, with maximum energy at 2$\omega$ of 220 J in 1 ns FWHM pulse duration.
It was focused by an f/3.5 lens to a 38 $\mu$m $1/e^{2}$ spot size, resulting in a laser intensity of $1 \times 10^{16}\mathrm{W/cm^{2}}$. 
The time delay between the main pulse and the pre-pulse in this case was 5 ns.\\

\setlength{\parindent}{0pt} 
\textbf{Diagnostics of the electron beams and the X-ray.}
The angular distribution of the generated electron beams was measured by a pair of image plates (IP, model Fuji- film BAS-SR 2040). 
They were also used to measure the beam charge \cite{tanaka2005}. 
There were copper filters with thickness 0.3-1 mm in front of each IP to provide the ability to measure the angular distribution over different energy ranges in a single shot. 

\setlength{\parindent}{12pt}
To avoid overestimating the beam charge due to the generation of hard X-ray photons when the electrons penetrated the metallic filters, simulations were performed using the Monte Carlo N-particle transport code (MCNP) \cite{Briesmeister2000} to calculate the average number of photons generated by each electron. 
The parameters of the target (filter and image plate) and the electron beam in the simulations were similar to the experiment. 
We found that the average number of photons generated is 0.32/electron.
The photostimulated luminescence (PSL) contribution from photons is only $\sim 1.6\%$ of the electron contribution due to a much smaller sensitivity of the IP to photons than to electrons, as shown in Ref. \cite{tanaka2005} for electrons and Ref. \cite{maddox2011} for photons. 
Therefore, the photons generated by electrons penetrating the filter can be neglected. 


\setlength{\parindent}{12pt}
The energy spectra of the electron beams were measured by a spectrometer with magnetic field strength of 9000 G and energy detection range of 0.9-49.4 MeV, which was placed behind the IPs. 
An X-ray pinhole camera with magnification M = 16 was used to measure the size of the plasma region. \\


\setlength{\parindent}{0pt}
\textbf{Simulations.}
The simulations were performed using the 2D3V PIC code EPOCH \cite{arber2015} on the TianHe supercomputer. 
The pulse duration of the incident laser is 270 fs (FWHM) with a spot size of $7~\mu$m. 
The wavelength, incident angle and polarization of the laser are the same as those in the experiment. 
The peak intensity of the laser is $2.8\times10^{20}\mathrm{W/cm^{2}}$.

\setlength{\parindent}{12pt}
The simulation box is initially located between $y = 30 ~\mu$m to $y = -140~\mu$m and $x = 0$ to $x = 150 ~\mu$m with a moving window in $x$ direction.
The target plasma is located between $y =-10~\mu$m and $y =-140~\mu$m with density profile of $n_{e}=10^{-(y+110)/25}n_{c}$ in $y$ direction, where $n_{c} = \epsilon_{0}m_{e}\omega_{0}^{2}/e^{2}$ is the critical density. 
The grid size is $\lambda_{L}/40$ in both directions and each cell contains 42 numerical macroparticles.
The density profile is given by the radiation hydrodynamic code MULTI \cite{Ramis2009977} by assuming that the contrast ratio of the pre-pulse with the main pulse is $10^{-6}$.

\setlength{\parindent}{12pt}
The work done by the electric field can be split into $x$, $y$, and $z$ components: 
\begin{equation}
\begin{aligned}
   W &= -\frac{e}{m_e c^2} \int_{0}^{t} \mathbf{E}.\mathbf{v} \, dt'  \\
       &= -\frac{e}{m_e c^2} \int_{0}^{t} (E_x v_x) + (E_y v_y) + (E_z v_z) \, dt'
\end{aligned}
\end{equation}
The EPOCH code was modified to track these components~\cite{shaw2017}, defined as:
\begin{equation}
   W_{i} = -\frac{e}{m_{e}c^{2}}\int_{0}^{t}E_{i}v_{i} \, dt', ~i = x,y,z
\end{equation}
This gives the energy gain caused by electric fields along each direction. 
In a 2D simulation with the laser polarized in $y$, the total energy gain can be written as $W = W_{x} + W_{y}$ since $E_{z} \equiv 0$.\\

\setlength{\parindent}{0pt}
\textbf{AUTHOR CONTRIBUTIONS.}
L. M. C. designed research; Y. M., L. M. C., J. R. Z., Y. F. L. and D. Z. L. performed experiment; S. J. D. D. modified the EPOCH source code. J. X. L., Y. M., Y. Y. M., X. H. Y., Z. Y. G performed the PIC simulations; Y. M., L. M. C., J. R. Z. and Y. F. L. analyzed data; Y. M., L. M. C. and Z. M. S. wrote the paper. J. Z. supported the whole project.\\

\setlength{\parindent}{0pt}
\textbf{ACKNOWLEDGMENTS.}
We thank Jupiter Laser Facility staffs at Lawrence Livermore National Laboratory (LLNL) for laser and technical support and Dr. Joseph Nilsen for facilitating the experiment. Y. M. would like to thank W. M. Wang, L. Willingale, A. G. R. Thomas, S. P. D. Mangles for fruitful discussions. This work was supported by the National Basic Research Program of China (2013CBA01500), National Key R$\&$D Program of China (2017YFA040330X),  National Key Scientific Instrument and Equipment Development Project (2012YQ120047), National Natural Science Foundation of China (11334013, 11721404), and NSAF (U1530150), the CAS key program (XDB17030500), the MOST International Collaboration (2014DFG02330), a Leverhulme Trust Research Project Grant, and the U.K. EPSRC grant EP/N028694/1. The EPOCH code was developed as part of the UK EPSRC grants EP/G056803/1 and EP/G055165/1.

\bibliography{Myrefs}

\begin{thebibliography}{51}%
\makeatletter
\providecommand \@ifxundefined [1]{%
 \@ifx{#1\undefined}
}%
\providecommand \@ifnum [1]{%
 \ifnum #1\expandafter \@firstoftwo
 \else \expandafter \@secondoftwo
 \fi
}%
\providecommand \@ifx [1]{%
 \ifx #1\expandafter \@firstoftwo
 \else \expandafter \@secondoftwo
 \fi
}%
\providecommand \natexlab [1]{#1}%
\providecommand \enquote  [1]{``#1''}%
\providecommand \bibnamefont  [1]{#1}%
\providecommand \bibfnamefont [1]{#1}%
\providecommand \citenamefont [1]{#1}%
\providecommand \href@noop [0]{\@secondoftwo}%
\providecommand \href [0]{\begingroup \@sanitize@url \@href}%
\providecommand \@href[1]{\@@startlink{#1}\@@href}%
\providecommand \@@href[1]{\endgroup#1\@@endlink}%
\providecommand \@sanitize@url [0]{\catcode `\\12\catcode `\$12\catcode
  `\&12\catcode `\#12\catcode `\^12\catcode `\_12\catcode `\%12\relax}%
\providecommand \@@startlink[1]{}%
\providecommand \@@endlink[0]{}%
\providecommand \url  [0]{\begingroup\@sanitize@url \@url }%
\providecommand \@url [1]{\endgroup\@href {#1}{\urlprefix }}%
\providecommand \urlprefix  [0]{URL }%
\providecommand \Eprint [0]{\href }%
\providecommand \doibase [0]{http://dx.doi.org/}%
\providecommand \selectlanguage [0]{\@gobble}%
\providecommand \bibinfo  [0]{\@secondoftwo}%
\providecommand \bibfield  [0]{\@secondoftwo}%
\providecommand \translation [1]{[#1]}%
\providecommand \BibitemOpen [0]{}%
\providecommand \bibitemStop [0]{}%
\providecommand \bibitemNoStop [0]{.\EOS\space}%
\providecommand \EOS [0]{\spacefactor3000\relax}%
\providecommand \BibitemShut  [1]{\csname bibitem#1\endcsname}%
\let\auto@bib@innerbib\@empty
\bibitem [{\citenamefont {Tajima}\ and\ \citenamefont
  {Dawson}(1979)}]{tajima1979}%
  \BibitemOpen
  \bibfield  {author} {\bibinfo {author} {\bibfnamefont {T.}~\bibnamefont
  {Tajima}}\ and\ \bibinfo {author} {\bibfnamefont {J.~M.}\ \bibnamefont
  {Dawson}},\ }\href {\doibase 10.1103/PhysRevLett.43.267} {\bibfield
  {journal} {\bibinfo  {journal} {Phys. Rev. Lett.}\ }\textbf {\bibinfo
  {volume} {43}},\ \bibinfo {pages} {267} (\bibinfo {year} {1979})}\BibitemShut
  {NoStop}%
\bibitem [{\citenamefont {Rosenbluth}\ and\ \citenamefont
  {Liu}(1972)}]{rosenbluth1972}%
  \BibitemOpen
  \bibfield  {author} {\bibinfo {author} {\bibfnamefont {M.~N.}\ \bibnamefont
  {Rosenbluth}}\ and\ \bibinfo {author} {\bibfnamefont {C.~S.}\ \bibnamefont
  {Liu}},\ }\href {\doibase 10.1103/PhysRevLett.29.701} {\bibfield  {journal}
  {\bibinfo  {journal} {Phys. Rev. Lett.}\ }\textbf {\bibinfo {volume} {29}},\
  \bibinfo {pages} {701} (\bibinfo {year} {1972})}\BibitemShut {NoStop}%
\bibitem [{\citenamefont {Krall}\ \emph {et~al.}(1993)\citenamefont {Krall},
  \citenamefont {Ting}, \citenamefont {Esarey},\ and\ \citenamefont
  {Sprangle}}]{krall1993}%
  \BibitemOpen
  \bibfield  {author} {\bibinfo {author} {\bibfnamefont {J.}~\bibnamefont
  {Krall}}, \bibinfo {author} {\bibfnamefont {A.}~\bibnamefont {Ting}},
  \bibinfo {author} {\bibfnamefont {E.}~\bibnamefont {Esarey}}, \ and\ \bibinfo
  {author} {\bibfnamefont {P.}~\bibnamefont {Sprangle}},\ }\href {\doibase
  10.1103/PhysRevE.48.2157} {\bibfield  {journal} {\bibinfo  {journal} {Phys.
  Rev. E}\ }\textbf {\bibinfo {volume} {48}},\ \bibinfo {pages} {2157}
  (\bibinfo {year} {1993})}\BibitemShut {NoStop}%
\bibitem [{\citenamefont {Sheng}\ \emph
  {et~al.}(2002{\natexlab{a}})\citenamefont {Sheng}, \citenamefont {Mima},
  \citenamefont {Sentoku}, \citenamefont {Nishihara},\ and\ \citenamefont
  {Zhang}}]{sheng2002}%
  \BibitemOpen
  \bibfield  {author} {\bibinfo {author} {\bibfnamefont {Z.}~\bibnamefont
  {Sheng}}, \bibinfo {author} {\bibfnamefont {K.}~\bibnamefont {Mima}},
  \bibinfo {author} {\bibfnamefont {Y.}~\bibnamefont {Sentoku}}, \bibinfo
  {author} {\bibfnamefont {K.}~\bibnamefont {Nishihara}}, \ and\ \bibinfo
  {author} {\bibfnamefont {J.}~\bibnamefont {Zhang}},\ }\href {\doibase
  10.1063/1.1485771} {\bibfield  {journal} {\bibinfo  {journal} {Physics of
  Plasmas}\ }\textbf {\bibinfo {volume} {9}},\ \bibinfo {pages} {3147}
  (\bibinfo {year} {2002}{\natexlab{a}})}\BibitemShut {NoStop}%
\bibitem [{\citenamefont {Pukhov}\ and\ \citenamefont {Meyer-ter
  Vehn}(2002)}]{pukhov2002}%
  \BibitemOpen
  \bibfield  {author} {\bibinfo {author} {\bibfnamefont {A.}~\bibnamefont
  {Pukhov}}\ and\ \bibinfo {author} {\bibfnamefont {J.}~\bibnamefont {Meyer-ter
  Vehn}},\ }\href {\doibase 10.1007/s003400200795} {\bibfield  {journal}
  {\bibinfo  {journal} {Applied Physics B}\ }\textbf {\bibinfo {volume} {74}},\
  \bibinfo {pages} {355} (\bibinfo {year} {2002})}\BibitemShut {NoStop}%
\bibitem [{\citenamefont {Kostyukov}\ \emph {et~al.}(2004)\citenamefont
  {Kostyukov}, \citenamefont {Pukhov},\ and\ \citenamefont
  {Kiselev}}]{kostyukov2004}%
  \BibitemOpen
  \bibfield  {author} {\bibinfo {author} {\bibfnamefont {I.}~\bibnamefont
  {Kostyukov}}, \bibinfo {author} {\bibfnamefont {A.}~\bibnamefont {Pukhov}}, \
  and\ \bibinfo {author} {\bibfnamefont {S.}~\bibnamefont {Kiselev}},\ }\href
  {\doibase 10.1063/1.1799371} {\bibfield  {journal} {\bibinfo  {journal}
  {Physics of Plasmas}\ }\textbf {\bibinfo {volume} {11}},\ \bibinfo {pages}
  {5256} (\bibinfo {year} {2004})}\BibitemShut {NoStop}%
\bibitem [{\citenamefont {Mangles}\ \emph {et~al.}(2004)\citenamefont
  {Mangles}, \citenamefont {Murphy}, \citenamefont {Najmudin}, \citenamefont
  {Thomas}, \citenamefont {Collier}, \citenamefont {Dangor}, \citenamefont
  {Divall}, \citenamefont {Foster}, \citenamefont {Gallacher}, \citenamefont
  {Hooker}, \citenamefont {Jaroszynski}, \citenamefont {Langley}, \citenamefont
  {Mori}, \citenamefont {Norreys}, \citenamefont {Tsung}, \citenamefont
  {Viskup}, \citenamefont {Walton},\ and\ \citenamefont
  {Krushelnick}}]{mangles2004}%
  \BibitemOpen
  \bibfield  {author} {\bibinfo {author} {\bibfnamefont {S.}~\bibnamefont
  {Mangles}}, \bibinfo {author} {\bibfnamefont {C.}~\bibnamefont {Murphy}},
  \bibinfo {author} {\bibfnamefont {Z.}~\bibnamefont {Najmudin}}, \bibinfo
  {author} {\bibfnamefont {A.}~\bibnamefont {Thomas}}, \bibinfo {author}
  {\bibfnamefont {J.}~\bibnamefont {Collier}}, \bibinfo {author} {\bibfnamefont
  {A.}~\bibnamefont {Dangor}}, \bibinfo {author} {\bibfnamefont
  {E.}~\bibnamefont {Divall}}, \bibinfo {author} {\bibfnamefont
  {P.}~\bibnamefont {Foster}}, \bibinfo {author} {\bibfnamefont
  {J.}~\bibnamefont {Gallacher}}, \bibinfo {author} {\bibfnamefont
  {C.}~\bibnamefont {Hooker}}, \bibinfo {author} {\bibfnamefont
  {D.}~\bibnamefont {Jaroszynski}}, \bibinfo {author} {\bibfnamefont
  {A.}~\bibnamefont {Langley}}, \bibinfo {author} {\bibfnamefont
  {W.}~\bibnamefont {Mori}}, \bibinfo {author} {\bibfnamefont {P.}~\bibnamefont
  {Norreys}}, \bibinfo {author} {\bibfnamefont {F.}~\bibnamefont {Tsung}},
  \bibinfo {author} {\bibfnamefont {R.}~\bibnamefont {Viskup}}, \bibinfo
  {author} {\bibfnamefont {B.}~\bibnamefont {Walton}}, \ and\ \bibinfo {author}
  {\bibfnamefont {K.}~\bibnamefont {Krushelnick}},\ }\href {\doibase
  10.1038/nature02939} {\bibfield  {journal} {\bibinfo  {journal} {Nature}\
  }\textbf {\bibinfo {volume} {431}},\ \bibinfo {pages} {535} (\bibinfo {year}
  {2004})}\BibitemShut {NoStop}%
\bibitem [{\citenamefont {Geddes}\ \emph {et~al.}(2004)\citenamefont {Geddes},
  \citenamefont {Toth}, \citenamefont {van Tilborg}, \citenamefont {Esarey},
  \citenamefont {Schroeder}, \citenamefont {Bruhwiler}, \citenamefont {Nieter},
  \citenamefont {Cary},\ and\ \citenamefont {Leemans}}]{geddes2004}%
  \BibitemOpen
  \bibfield  {author} {\bibinfo {author} {\bibfnamefont {C.}~\bibnamefont
  {Geddes}}, \bibinfo {author} {\bibfnamefont {C.}~\bibnamefont {Toth}},
  \bibinfo {author} {\bibfnamefont {J.}~\bibnamefont {van Tilborg}}, \bibinfo
  {author} {\bibfnamefont {E.}~\bibnamefont {Esarey}}, \bibinfo {author}
  {\bibfnamefont {C.}~\bibnamefont {Schroeder}}, \bibinfo {author}
  {\bibfnamefont {D.}~\bibnamefont {Bruhwiler}}, \bibinfo {author}
  {\bibfnamefont {C.}~\bibnamefont {Nieter}}, \bibinfo {author} {\bibfnamefont
  {J.}~\bibnamefont {Cary}}, \ and\ \bibinfo {author} {\bibfnamefont
  {W.}~\bibnamefont {Leemans}},\ }\href {\doibase 10.1038/nature02900}
  {\bibfield  {journal} {\bibinfo  {journal} {Nature}\ }\textbf {\bibinfo
  {volume} {431}},\ \bibinfo {pages} {538} (\bibinfo {year}
  {2004})}\BibitemShut {NoStop}%
\bibitem [{\citenamefont {Faure}\ \emph {et~al.}(2004)\citenamefont {Faure},
  \citenamefont {Glinec}, \citenamefont {Pukhov}, \citenamefont {Kiselev},
  \citenamefont {Gordienko}, \citenamefont {Lefebvre}, \citenamefont
  {Rousseau}, \citenamefont {Burgy},\ and\ \citenamefont {Malka}}]{faure2004}%
  \BibitemOpen
  \bibfield  {author} {\bibinfo {author} {\bibfnamefont {J.}~\bibnamefont
  {Faure}}, \bibinfo {author} {\bibfnamefont {Y.}~\bibnamefont {Glinec}},
  \bibinfo {author} {\bibfnamefont {A.}~\bibnamefont {Pukhov}}, \bibinfo
  {author} {\bibfnamefont {S.}~\bibnamefont {Kiselev}}, \bibinfo {author}
  {\bibfnamefont {S.}~\bibnamefont {Gordienko}}, \bibinfo {author}
  {\bibfnamefont {E.}~\bibnamefont {Lefebvre}}, \bibinfo {author}
  {\bibfnamefont {J.}~\bibnamefont {Rousseau}}, \bibinfo {author}
  {\bibfnamefont {F.}~\bibnamefont {Burgy}}, \ and\ \bibinfo {author}
  {\bibfnamefont {V.}~\bibnamefont {Malka}},\ }\href {\doibase
  10.1038/nature02963} {\bibfield  {journal} {\bibinfo  {journal} {Nature}\
  }\textbf {\bibinfo {volume} {431}},\ \bibinfo {pages} {541} (\bibinfo {year}
  {2004})}\BibitemShut {NoStop}%
\bibitem [{\citenamefont {Leemans}\ \emph {et~al.}(2006)\citenamefont
  {Leemans}, \citenamefont {Nagler}, \citenamefont {Gonsalves}, \citenamefont
  {Toth}, \citenamefont {Nakamura}, \citenamefont {Geddes}, \citenamefont
  {Esarey}, \citenamefont {Schroeder},\ and\ \citenamefont
  {Hooker}}]{leemans2006}%
  \BibitemOpen
  \bibfield  {author} {\bibinfo {author} {\bibfnamefont {W.~P.}\ \bibnamefont
  {Leemans}}, \bibinfo {author} {\bibfnamefont {B.}~\bibnamefont {Nagler}},
  \bibinfo {author} {\bibfnamefont {A.~J.}\ \bibnamefont {Gonsalves}}, \bibinfo
  {author} {\bibfnamefont {C.}~\bibnamefont {Toth}}, \bibinfo {author}
  {\bibfnamefont {K.}~\bibnamefont {Nakamura}}, \bibinfo {author}
  {\bibfnamefont {C.~G.~R.}\ \bibnamefont {Geddes}}, \bibinfo {author}
  {\bibfnamefont {E.}~\bibnamefont {Esarey}}, \bibinfo {author} {\bibfnamefont
  {C.~B.}\ \bibnamefont {Schroeder}}, \ and\ \bibinfo {author} {\bibfnamefont
  {S.~M.}\ \bibnamefont {Hooker}},\ }\href {\doibase 10.1038/nphys418}
  {\bibfield  {journal} {\bibinfo  {journal} {Nature Physics}\ }\textbf
  {\bibinfo {volume} {2}},\ \bibinfo {pages} {696} (\bibinfo {year}
  {2006})}\BibitemShut {NoStop}%
\bibitem [{\citenamefont {Hafz}\ \emph {et~al.}(2008)\citenamefont {Hafz},
  \citenamefont {Jeong}, \citenamefont {Choi}, \citenamefont {Lee},
  \citenamefont {Pae}, \citenamefont {Kulagin}, \citenamefont {Sung},
  \citenamefont {Yu}, \citenamefont {Hong}, \citenamefont {Hosokai},
  \citenamefont {Cary}, \citenamefont {Ko},\ and\ \citenamefont
  {Lee}}]{hafz2008}%
  \BibitemOpen
  \bibfield  {author} {\bibinfo {author} {\bibfnamefont {N.~A.~M.}\
  \bibnamefont {Hafz}}, \bibinfo {author} {\bibfnamefont {T.~M.}\ \bibnamefont
  {Jeong}}, \bibinfo {author} {\bibfnamefont {I.~W.}\ \bibnamefont {Choi}},
  \bibinfo {author} {\bibfnamefont {S.~K.}\ \bibnamefont {Lee}}, \bibinfo
  {author} {\bibfnamefont {K.~H.}\ \bibnamefont {Pae}}, \bibinfo {author}
  {\bibfnamefont {V.~V.}\ \bibnamefont {Kulagin}}, \bibinfo {author}
  {\bibfnamefont {J.~H.}\ \bibnamefont {Sung}}, \bibinfo {author}
  {\bibfnamefont {T.~J.}\ \bibnamefont {Yu}}, \bibinfo {author} {\bibfnamefont
  {K.-H.}\ \bibnamefont {Hong}}, \bibinfo {author} {\bibfnamefont
  {T.}~\bibnamefont {Hosokai}}, \bibinfo {author} {\bibfnamefont {J.~R.}\
  \bibnamefont {Cary}}, \bibinfo {author} {\bibfnamefont {D.-K.}\ \bibnamefont
  {Ko}}, \ and\ \bibinfo {author} {\bibfnamefont {J.}~\bibnamefont {Lee}},\
  }\href {\doibase 10.1038/nphoton.2008.155} {\bibfield  {journal} {\bibinfo
  {journal} {Nature Photonics}\ }\textbf {\bibinfo {volume} {2}},\ \bibinfo
  {pages} {571} (\bibinfo {year} {2008})}\BibitemShut {NoStop}%
\bibitem [{\citenamefont {Wang}\ \emph
  {et~al.}(2013{\natexlab{a}})\citenamefont {Wang}, \citenamefont {Zgadzaj},
  \citenamefont {Fazel}, \citenamefont {Li}, \citenamefont {Yi}, \citenamefont
  {Zhang}, \citenamefont {Henderson}, \citenamefont {Chang}, \citenamefont
  {Korzekwa}, \citenamefont {Tsai}, \citenamefont {Pai}, \citenamefont
  {Quevedo}, \citenamefont {Dyer}, \citenamefont {Gaul}, \citenamefont
  {Martinez}, \citenamefont {Bernstein}, \citenamefont {Borger}, \citenamefont
  {Spinks}, \citenamefont {Donovan}, \citenamefont {Khudik}, \citenamefont
  {Shvets}, \citenamefont {Ditmire},\ and\ \citenamefont {Downer}}]{wang2013}%
  \BibitemOpen
  \bibfield  {author} {\bibinfo {author} {\bibfnamefont {X.}~\bibnamefont
  {Wang}}, \bibinfo {author} {\bibfnamefont {R.}~\bibnamefont {Zgadzaj}},
  \bibinfo {author} {\bibfnamefont {N.}~\bibnamefont {Fazel}}, \bibinfo
  {author} {\bibfnamefont {Z.}~\bibnamefont {Li}}, \bibinfo {author}
  {\bibfnamefont {S.~A.}\ \bibnamefont {Yi}}, \bibinfo {author} {\bibfnamefont
  {X.}~\bibnamefont {Zhang}}, \bibinfo {author} {\bibfnamefont
  {W.}~\bibnamefont {Henderson}}, \bibinfo {author} {\bibfnamefont {Y.~Y.}\
  \bibnamefont {Chang}}, \bibinfo {author} {\bibfnamefont {R.}~\bibnamefont
  {Korzekwa}}, \bibinfo {author} {\bibfnamefont {H.~E.}\ \bibnamefont {Tsai}},
  \bibinfo {author} {\bibfnamefont {C.~H.}\ \bibnamefont {Pai}}, \bibinfo
  {author} {\bibfnamefont {H.}~\bibnamefont {Quevedo}}, \bibinfo {author}
  {\bibfnamefont {G.}~\bibnamefont {Dyer}}, \bibinfo {author} {\bibfnamefont
  {E.}~\bibnamefont {Gaul}}, \bibinfo {author} {\bibfnamefont {M.}~\bibnamefont
  {Martinez}}, \bibinfo {author} {\bibfnamefont {A.~C.}\ \bibnamefont
  {Bernstein}}, \bibinfo {author} {\bibfnamefont {T.}~\bibnamefont {Borger}},
  \bibinfo {author} {\bibfnamefont {M.}~\bibnamefont {Spinks}}, \bibinfo
  {author} {\bibfnamefont {M.}~\bibnamefont {Donovan}}, \bibinfo {author}
  {\bibfnamefont {V.}~\bibnamefont {Khudik}}, \bibinfo {author} {\bibfnamefont
  {G.}~\bibnamefont {Shvets}}, \bibinfo {author} {\bibfnamefont
  {T.}~\bibnamefont {Ditmire}}, \ and\ \bibinfo {author} {\bibfnamefont
  {M.~C.}\ \bibnamefont {Downer}},\ }\href {\doibase 10.1038/ncomms2988}
  {\bibfield  {journal} {\bibinfo  {journal} {Nature Communications}\ }\textbf
  {\bibinfo {volume} {4}},\ \bibinfo {pages} {1988} (\bibinfo {year}
  {2013}{\natexlab{a}})}\BibitemShut {NoStop}%
\bibitem [{\citenamefont {Kim}\ \emph {et~al.}(2013)\citenamefont {Kim},
  \citenamefont {Pae}, \citenamefont {Cha}, \citenamefont {Kim}, \citenamefont
  {Yu}, \citenamefont {Sung}, \citenamefont {Lee}, \citenamefont {Jeong},\ and\
  \citenamefont {Lee}}]{kim2013}%
  \BibitemOpen
  \bibfield  {author} {\bibinfo {author} {\bibfnamefont {H.~T.}\ \bibnamefont
  {Kim}}, \bibinfo {author} {\bibfnamefont {K.~H.}\ \bibnamefont {Pae}},
  \bibinfo {author} {\bibfnamefont {H.~J.}\ \bibnamefont {Cha}}, \bibinfo
  {author} {\bibfnamefont {I.~J.}\ \bibnamefont {Kim}}, \bibinfo {author}
  {\bibfnamefont {T.~J.}\ \bibnamefont {Yu}}, \bibinfo {author} {\bibfnamefont
  {J.~H.}\ \bibnamefont {Sung}}, \bibinfo {author} {\bibfnamefont {S.~K.}\
  \bibnamefont {Lee}}, \bibinfo {author} {\bibfnamefont {T.~M.}\ \bibnamefont
  {Jeong}}, \ and\ \bibinfo {author} {\bibfnamefont {J.}~\bibnamefont {Lee}},\
  }\href {\doibase 10.1103/PhysRevLett.111.165002} {\bibfield  {journal}
  {\bibinfo  {journal} {Phys. Rev. Lett.}\ }\textbf {\bibinfo {volume} {111}},\
  \bibinfo {pages} {165002} (\bibinfo {year} {2013})}\BibitemShut {NoStop}%
\bibitem [{\citenamefont {Leemans}\ \emph {et~al.}(2014)\citenamefont
  {Leemans}, \citenamefont {Gonsalves}, \citenamefont {Mao}, \citenamefont
  {Nakamura}, \citenamefont {Benedetti}, \citenamefont {Schroeder},
  \citenamefont {T\'oth}, \citenamefont {Daniels}, \citenamefont
  {Mittelberger}, \citenamefont {Bulanov}, \citenamefont {Vay}, \citenamefont
  {Geddes},\ and\ \citenamefont {Esarey}}]{leemans2014}%
  \BibitemOpen
  \bibfield  {author} {\bibinfo {author} {\bibfnamefont {W.~P.}\ \bibnamefont
  {Leemans}}, \bibinfo {author} {\bibfnamefont {A.~J.}\ \bibnamefont
  {Gonsalves}}, \bibinfo {author} {\bibfnamefont {H.-S.}\ \bibnamefont {Mao}},
  \bibinfo {author} {\bibfnamefont {K.}~\bibnamefont {Nakamura}}, \bibinfo
  {author} {\bibfnamefont {C.}~\bibnamefont {Benedetti}}, \bibinfo {author}
  {\bibfnamefont {C.~B.}\ \bibnamefont {Schroeder}}, \bibinfo {author}
  {\bibfnamefont {C.}~\bibnamefont {T\'oth}}, \bibinfo {author} {\bibfnamefont
  {J.}~\bibnamefont {Daniels}}, \bibinfo {author} {\bibfnamefont {D.~E.}\
  \bibnamefont {Mittelberger}}, \bibinfo {author} {\bibfnamefont {S.~S.}\
  \bibnamefont {Bulanov}}, \bibinfo {author} {\bibfnamefont {J.-L.}\
  \bibnamefont {Vay}}, \bibinfo {author} {\bibfnamefont {C.~G.~R.}\
  \bibnamefont {Geddes}}, \ and\ \bibinfo {author} {\bibfnamefont
  {E.}~\bibnamefont {Esarey}},\ }\href {\doibase
  10.1103/PhysRevLett.113.245002} {\bibfield  {journal} {\bibinfo  {journal}
  {Phys. Rev. Lett.}\ }\textbf {\bibinfo {volume} {113}},\ \bibinfo {pages}
  {245002} (\bibinfo {year} {2014})}\BibitemShut {NoStop}%
\bibitem [{\citenamefont {Chen}\ \emph
  {et~al.}(2001{\natexlab{a}})\citenamefont {Chen}, \citenamefont {Zhang},
  \citenamefont {Li}, \citenamefont {Teng}, \citenamefont {Liang},
  \citenamefont {Sheng}, \citenamefont {Dong}, \citenamefont {Zhao},
  \citenamefont {Wei},\ and\ \citenamefont {Tang}}]{chenlm2001}%
  \BibitemOpen
  \bibfield  {author} {\bibinfo {author} {\bibfnamefont {L.~M.}\ \bibnamefont
  {Chen}}, \bibinfo {author} {\bibfnamefont {J.}~\bibnamefont {Zhang}},
  \bibinfo {author} {\bibfnamefont {Y.~T.}\ \bibnamefont {Li}}, \bibinfo
  {author} {\bibfnamefont {H.}~\bibnamefont {Teng}}, \bibinfo {author}
  {\bibfnamefont {T.~J.}\ \bibnamefont {Liang}}, \bibinfo {author}
  {\bibfnamefont {Z.~M.}\ \bibnamefont {Sheng}}, \bibinfo {author}
  {\bibfnamefont {Q.~L.}\ \bibnamefont {Dong}}, \bibinfo {author}
  {\bibfnamefont {L.~Z.}\ \bibnamefont {Zhao}}, \bibinfo {author}
  {\bibfnamefont {Z.~Y.}\ \bibnamefont {Wei}}, \ and\ \bibinfo {author}
  {\bibfnamefont {X.~W.}\ \bibnamefont {Tang}},\ }\href {\doibase
  10.1103/PhysRevLett.87.225001} {\bibfield  {journal} {\bibinfo  {journal}
  {Phys. Rev. Lett.}\ }\textbf {\bibinfo {volume} {87}},\ \bibinfo {pages}
  {225001} (\bibinfo {year} {2001}{\natexlab{a}})}\BibitemShut {NoStop}%
\bibitem [{\citenamefont {Li}\ \emph {et~al.}(2001)\citenamefont {Li},
  \citenamefont {Zhang}, \citenamefont {Chen}, \citenamefont {Mu},
  \citenamefont {Liang}, \citenamefont {Wei}, \citenamefont {Dong},
  \citenamefont {Chen}, \citenamefont {Teng}, \citenamefont {Chun-Yu},
  \citenamefont {Jiang}, \citenamefont {Zheng},\ and\ \citenamefont
  {Tang}}]{liyt2001}%
  \BibitemOpen
  \bibfield  {author} {\bibinfo {author} {\bibfnamefont {Y.~T.}\ \bibnamefont
  {Li}}, \bibinfo {author} {\bibfnamefont {J.}~\bibnamefont {Zhang}}, \bibinfo
  {author} {\bibfnamefont {L.~M.}\ \bibnamefont {Chen}}, \bibinfo {author}
  {\bibfnamefont {Y.~F.}\ \bibnamefont {Mu}}, \bibinfo {author} {\bibfnamefont
  {T.~J.}\ \bibnamefont {Liang}}, \bibinfo {author} {\bibfnamefont {Z.~Y.}\
  \bibnamefont {Wei}}, \bibinfo {author} {\bibfnamefont {Q.~L.}\ \bibnamefont
  {Dong}}, \bibinfo {author} {\bibfnamefont {Z.~L.}\ \bibnamefont {Chen}},
  \bibinfo {author} {\bibfnamefont {H.}~\bibnamefont {Teng}}, \bibinfo {author}
  {\bibfnamefont {S.~T.}\ \bibnamefont {Chun-Yu}}, \bibinfo {author}
  {\bibfnamefont {W.~M.}\ \bibnamefont {Jiang}}, \bibinfo {author}
  {\bibfnamefont {Z.~J.}\ \bibnamefont {Zheng}}, \ and\ \bibinfo {author}
  {\bibfnamefont {X.~W.}\ \bibnamefont {Tang}},\ }\href {\doibase
  10.1103/PhysRevE.64.046407} {\bibfield  {journal} {\bibinfo  {journal} {Phys.
  Rev. E}\ }\textbf {\bibinfo {volume} {64}},\ \bibinfo {pages} {046407}
  (\bibinfo {year} {2001})}\BibitemShut {NoStop}%
\bibitem [{\citenamefont {Li}\ \emph {et~al.}(2006)\citenamefont {Li},
  \citenamefont {Yuan}, \citenamefont {Xu}, \citenamefont {Zheng},
  \citenamefont {Sheng}, \citenamefont {Chen}, \citenamefont {Ma},
  \citenamefont {Liang}, \citenamefont {Yu}, \citenamefont {Zhang},
  \citenamefont {Liu}, \citenamefont {Wang}, \citenamefont {Wei}, \citenamefont
  {Zhao}, \citenamefont {Jin},\ and\ \citenamefont {Zhang}}]{liyt2006}%
  \BibitemOpen
  \bibfield  {author} {\bibinfo {author} {\bibfnamefont {Y.~T.}\ \bibnamefont
  {Li}}, \bibinfo {author} {\bibfnamefont {X.~H.}\ \bibnamefont {Yuan}},
  \bibinfo {author} {\bibfnamefont {M.~H.}\ \bibnamefont {Xu}}, \bibinfo
  {author} {\bibfnamefont {Z.~Y.}\ \bibnamefont {Zheng}}, \bibinfo {author}
  {\bibfnamefont {Z.~M.}\ \bibnamefont {Sheng}}, \bibinfo {author}
  {\bibfnamefont {M.}~\bibnamefont {Chen}}, \bibinfo {author} {\bibfnamefont
  {Y.~Y.}\ \bibnamefont {Ma}}, \bibinfo {author} {\bibfnamefont {W.~X.}\
  \bibnamefont {Liang}}, \bibinfo {author} {\bibfnamefont {Q.~Z.}\ \bibnamefont
  {Yu}}, \bibinfo {author} {\bibfnamefont {Y.}~\bibnamefont {Zhang}}, \bibinfo
  {author} {\bibfnamefont {F.}~\bibnamefont {Liu}}, \bibinfo {author}
  {\bibfnamefont {Z.~H.}\ \bibnamefont {Wang}}, \bibinfo {author}
  {\bibfnamefont {Z.~Y.}\ \bibnamefont {Wei}}, \bibinfo {author} {\bibfnamefont
  {W.}~\bibnamefont {Zhao}}, \bibinfo {author} {\bibfnamefont {Z.}~\bibnamefont
  {Jin}}, \ and\ \bibinfo {author} {\bibfnamefont {J.}~\bibnamefont {Zhang}},\
  }\href {\doibase 10.1103/PhysRevLett.96.165003} {\bibfield  {journal}
  {\bibinfo  {journal} {Phys. Rev. Lett.}\ }\textbf {\bibinfo {volume} {96}},\
  \bibinfo {pages} {165003} (\bibinfo {year} {2006})}\BibitemShut {NoStop}%
\bibitem [{\citenamefont {Habara}\ \emph {et~al.}(2006)\citenamefont {Habara},
  \citenamefont {Adumi}, \citenamefont {Yabuuchi}, \citenamefont {Nakamura},
  \citenamefont {Chen}, \citenamefont {Kashihara}, \citenamefont {Kodama},
  \citenamefont {Kondo}, \citenamefont {Kumar}, \citenamefont {Lei},
  \citenamefont {Matsuoka}, \citenamefont {Mima},\ and\ \citenamefont
  {Tanaka}}]{habara2006}%
  \BibitemOpen
  \bibfield  {author} {\bibinfo {author} {\bibfnamefont {H.}~\bibnamefont
  {Habara}}, \bibinfo {author} {\bibfnamefont {K.}~\bibnamefont {Adumi}},
  \bibinfo {author} {\bibfnamefont {T.}~\bibnamefont {Yabuuchi}}, \bibinfo
  {author} {\bibfnamefont {T.}~\bibnamefont {Nakamura}}, \bibinfo {author}
  {\bibfnamefont {Z.~L.}\ \bibnamefont {Chen}}, \bibinfo {author}
  {\bibfnamefont {M.}~\bibnamefont {Kashihara}}, \bibinfo {author}
  {\bibfnamefont {R.}~\bibnamefont {Kodama}}, \bibinfo {author} {\bibfnamefont
  {K.}~\bibnamefont {Kondo}}, \bibinfo {author} {\bibfnamefont {G.~R.}\
  \bibnamefont {Kumar}}, \bibinfo {author} {\bibfnamefont {L.~A.}\ \bibnamefont
  {Lei}}, \bibinfo {author} {\bibfnamefont {T.}~\bibnamefont {Matsuoka}},
  \bibinfo {author} {\bibfnamefont {K.}~\bibnamefont {Mima}}, \ and\ \bibinfo
  {author} {\bibfnamefont {K.~A.}\ \bibnamefont {Tanaka}},\ }\href {\doibase
  10.1103/PhysRevLett.97.095004} {\bibfield  {journal} {\bibinfo  {journal}
  {Phys. Rev. Lett.}\ }\textbf {\bibinfo {volume} {97}},\ \bibinfo {pages}
  {095004} (\bibinfo {year} {2006})}\BibitemShut {NoStop}%
\bibitem [{\citenamefont {Mordovanakis}\ \emph {et~al.}(2009)\citenamefont
  {Mordovanakis}, \citenamefont {Easter}, \citenamefont {Naumova},
  \citenamefont {Popov}, \citenamefont {Masson-Laborde}, \citenamefont {Hou},
  \citenamefont {Sokolov}, \citenamefont {Mourou}, \citenamefont {Glazyrin},
  \citenamefont {Rozmus}, \citenamefont {Bychenkov}, \citenamefont {Nees},\
  and\ \citenamefont {Krushelnick}}]{mordo2009}%
  \BibitemOpen
  \bibfield  {author} {\bibinfo {author} {\bibfnamefont {A.~G.}\ \bibnamefont
  {Mordovanakis}}, \bibinfo {author} {\bibfnamefont {J.}~\bibnamefont
  {Easter}}, \bibinfo {author} {\bibfnamefont {N.}~\bibnamefont {Naumova}},
  \bibinfo {author} {\bibfnamefont {K.}~\bibnamefont {Popov}}, \bibinfo
  {author} {\bibfnamefont {P.-E.}\ \bibnamefont {Masson-Laborde}}, \bibinfo
  {author} {\bibfnamefont {B.}~\bibnamefont {Hou}}, \bibinfo {author}
  {\bibfnamefont {I.}~\bibnamefont {Sokolov}}, \bibinfo {author} {\bibfnamefont
  {G.}~\bibnamefont {Mourou}}, \bibinfo {author} {\bibfnamefont {I.~V.}\
  \bibnamefont {Glazyrin}}, \bibinfo {author} {\bibfnamefont {W.}~\bibnamefont
  {Rozmus}}, \bibinfo {author} {\bibfnamefont {V.}~\bibnamefont {Bychenkov}},
  \bibinfo {author} {\bibfnamefont {J.}~\bibnamefont {Nees}}, \ and\ \bibinfo
  {author} {\bibfnamefont {K.}~\bibnamefont {Krushelnick}},\ }\href {\doibase
  10.1103/PhysRevLett.103.235001} {\bibfield  {journal} {\bibinfo  {journal}
  {Phys. Rev. Lett.}\ }\textbf {\bibinfo {volume} {103}},\ \bibinfo {pages}
  {235001} (\bibinfo {year} {2009})}\BibitemShut {NoStop}%
\bibitem [{\citenamefont {Tian}\ \emph {et~al.}(2012)\citenamefont {Tian},
  \citenamefont {Liu}, \citenamefont {Wang}, \citenamefont {Wang},
  \citenamefont {Deng}, \citenamefont {Xia}, \citenamefont {Li}, \citenamefont
  {Cao}, \citenamefont {Lu}, \citenamefont {Zhang}, \citenamefont {Xu},
  \citenamefont {Leng}, \citenamefont {Li},\ and\ \citenamefont
  {Xu}}]{tian2012}%
  \BibitemOpen
  \bibfield  {author} {\bibinfo {author} {\bibfnamefont {Y.}~\bibnamefont
  {Tian}}, \bibinfo {author} {\bibfnamefont {J.}~\bibnamefont {Liu}}, \bibinfo
  {author} {\bibfnamefont {W.}~\bibnamefont {Wang}}, \bibinfo {author}
  {\bibfnamefont {C.}~\bibnamefont {Wang}}, \bibinfo {author} {\bibfnamefont
  {A.}~\bibnamefont {Deng}}, \bibinfo {author} {\bibfnamefont {C.}~\bibnamefont
  {Xia}}, \bibinfo {author} {\bibfnamefont {W.}~\bibnamefont {Li}}, \bibinfo
  {author} {\bibfnamefont {L.}~\bibnamefont {Cao}}, \bibinfo {author}
  {\bibfnamefont {H.}~\bibnamefont {Lu}}, \bibinfo {author} {\bibfnamefont
  {H.}~\bibnamefont {Zhang}}, \bibinfo {author} {\bibfnamefont
  {Y.}~\bibnamefont {Xu}}, \bibinfo {author} {\bibfnamefont {Y.}~\bibnamefont
  {Leng}}, \bibinfo {author} {\bibfnamefont {R.}~\bibnamefont {Li}}, \ and\
  \bibinfo {author} {\bibfnamefont {Z.}~\bibnamefont {Xu}},\ }\href {\doibase
  10.1103/PhysRevLett.109.115002} {\bibfield  {journal} {\bibinfo  {journal}
  {Phys. Rev. Lett.}\ }\textbf {\bibinfo {volume} {109}},\ \bibinfo {pages}
  {115002} (\bibinfo {year} {2012})}\BibitemShut {NoStop}%
\bibitem [{\citenamefont {Mao}\ \emph {et~al.}(2012)\citenamefont {Mao},
  \citenamefont {Chen}, \citenamefont {Ge}, \citenamefont {Zhang},
  \citenamefont {Yan}, \citenamefont {Li}, \citenamefont {Liao}, \citenamefont
  {Ma}, \citenamefont {Huang}, \citenamefont {Li}, \citenamefont {Lu},
  \citenamefont {Dong}, \citenamefont {Wei}, \citenamefont {Sheng},\ and\
  \citenamefont {Zhang}}]{mao2012}%
  \BibitemOpen
  \bibfield  {author} {\bibinfo {author} {\bibfnamefont {J.~Y.}\ \bibnamefont
  {Mao}}, \bibinfo {author} {\bibfnamefont {L.~M.}\ \bibnamefont {Chen}},
  \bibinfo {author} {\bibfnamefont {X.~L.}\ \bibnamefont {Ge}}, \bibinfo
  {author} {\bibfnamefont {L.}~\bibnamefont {Zhang}}, \bibinfo {author}
  {\bibfnamefont {W.~C.}\ \bibnamefont {Yan}}, \bibinfo {author} {\bibfnamefont
  {D.~Z.}\ \bibnamefont {Li}}, \bibinfo {author} {\bibfnamefont {G.~Q.}\
  \bibnamefont {Liao}}, \bibinfo {author} {\bibfnamefont {J.~L.}\ \bibnamefont
  {Ma}}, \bibinfo {author} {\bibfnamefont {K.}~\bibnamefont {Huang}}, \bibinfo
  {author} {\bibfnamefont {Y.~T.}\ \bibnamefont {Li}}, \bibinfo {author}
  {\bibfnamefont {X.}~\bibnamefont {Lu}}, \bibinfo {author} {\bibfnamefont
  {Q.~L.}\ \bibnamefont {Dong}}, \bibinfo {author} {\bibfnamefont {Z.~Y.}\
  \bibnamefont {Wei}}, \bibinfo {author} {\bibfnamefont {Z.~M.}\ \bibnamefont
  {Sheng}}, \ and\ \bibinfo {author} {\bibfnamefont {J.}~\bibnamefont
  {Zhang}},\ }\href {\doibase 10.1103/PhysRevE.85.025401} {\bibfield  {journal}
  {\bibinfo  {journal} {Phys. Rev. E}\ }\textbf {\bibinfo {volume} {85}},\
  \bibinfo {pages} {025401} (\bibinfo {year} {2012})}\BibitemShut {NoStop}%
\bibitem [{\citenamefont {Wang}\ \emph
  {et~al.}(2013{\natexlab{b}})\citenamefont {Wang}, \citenamefont {Chen},
  \citenamefont {Mao}, \citenamefont {Huang}, \citenamefont {Ma}, \citenamefont
  {Zhao}, \citenamefont {Zhang}, \citenamefont {Yan}, \citenamefont {Li},
  \citenamefont {Ma}, \citenamefont {Li}, \citenamefont {Lu}, \citenamefont
  {Wei}, \citenamefont {Sheng},\ and\ \citenamefont {Zhang}}]{Wang2013578}%
  \BibitemOpen
  \bibfield  {author} {\bibinfo {author} {\bibfnamefont {W.}~\bibnamefont
  {Wang}}, \bibinfo {author} {\bibfnamefont {L.}~\bibnamefont {Chen}}, \bibinfo
  {author} {\bibfnamefont {J.}~\bibnamefont {Mao}}, \bibinfo {author}
  {\bibfnamefont {K.}~\bibnamefont {Huang}}, \bibinfo {author} {\bibfnamefont
  {Y.}~\bibnamefont {Ma}}, \bibinfo {author} {\bibfnamefont {J.}~\bibnamefont
  {Zhao}}, \bibinfo {author} {\bibfnamefont {L.}~\bibnamefont {Zhang}},
  \bibinfo {author} {\bibfnamefont {W.}~\bibnamefont {Yan}}, \bibinfo {author}
  {\bibfnamefont {D.}~\bibnamefont {Li}}, \bibinfo {author} {\bibfnamefont
  {J.}~\bibnamefont {Ma}}, \bibinfo {author} {\bibfnamefont {Y.}~\bibnamefont
  {Li}}, \bibinfo {author} {\bibfnamefont {X.}~\bibnamefont {Lu}}, \bibinfo
  {author} {\bibfnamefont {Z.}~\bibnamefont {Wei}}, \bibinfo {author}
  {\bibfnamefont {Z.}~\bibnamefont {Sheng}}, \ and\ \bibinfo {author}
  {\bibfnamefont {J.}~\bibnamefont {Zhang}},\ }\href {\doibase
  http://dx.doi.org/10.1016/j.hedp.2013.05.011} {\bibfield  {journal} {\bibinfo
   {journal} {High Energy Density Physics}\ }\textbf {\bibinfo {volume} {9}},\
  \bibinfo {pages} {578 } (\bibinfo {year} {2013}{\natexlab{b}})}\BibitemShut
  {NoStop}%
\bibitem [{\citenamefont {Estabrook}\ and\ \citenamefont
  {Kruer}(1978)}]{estabrook1978}%
  \BibitemOpen
  \bibfield  {author} {\bibinfo {author} {\bibfnamefont {K.}~\bibnamefont
  {Estabrook}}\ and\ \bibinfo {author} {\bibfnamefont {W.~L.}\ \bibnamefont
  {Kruer}},\ }\href {\doibase 10.1103/PhysRevLett.40.42} {\bibfield  {journal}
  {\bibinfo  {journal} {Phys. Rev. Lett.}\ }\textbf {\bibinfo {volume} {40}},\
  \bibinfo {pages} {42} (\bibinfo {year} {1978})}\BibitemShut {NoStop}%
\bibitem [{\citenamefont {Brunel}(1987)}]{brunel1987}%
  \BibitemOpen
  \bibfield  {author} {\bibinfo {author} {\bibfnamefont {F.}~\bibnamefont
  {Brunel}},\ }\href {\doibase 10.1103/PhysRevLett.59.52} {\bibfield  {journal}
  {\bibinfo  {journal} {Phys. Rev. Lett.}\ }\textbf {\bibinfo {volume} {59}},\
  \bibinfo {pages} {52} (\bibinfo {year} {1987})}\BibitemShut {NoStop}%
\bibitem [{\citenamefont {Chen}\ \emph
  {et~al.}(2001{\natexlab{b}})\citenamefont {Chen}, \citenamefont {Zhang},
  \citenamefont {Dong}, \citenamefont {Teng}, \citenamefont {Liang},
  \citenamefont {Zhao},\ and\ \citenamefont {Wei}}]{chenlm2001-2}%
  \BibitemOpen
  \bibfield  {author} {\bibinfo {author} {\bibfnamefont {L.}~\bibnamefont
  {Chen}}, \bibinfo {author} {\bibfnamefont {J.}~\bibnamefont {Zhang}},
  \bibinfo {author} {\bibfnamefont {Q.}~\bibnamefont {Dong}}, \bibinfo {author}
  {\bibfnamefont {H.}~\bibnamefont {Teng}}, \bibinfo {author} {\bibfnamefont
  {T.}~\bibnamefont {Liang}}, \bibinfo {author} {\bibfnamefont
  {L.}~\bibnamefont {Zhao}}, \ and\ \bibinfo {author} {\bibfnamefont
  {Z.}~\bibnamefont {Wei}},\ }\href {\doibase 10.1063/1.1371956} {\bibfield
  {journal} {\bibinfo  {journal} {Physics of Plasmas}\ }\textbf {\bibinfo
  {volume} {8}},\ \bibinfo {pages} {2925} (\bibinfo {year}
  {2001}{\natexlab{b}})}\BibitemShut {NoStop}%
\bibitem [{\citenamefont {Kruer}\ and\ \citenamefont
  {Estabrook}(1985)}]{kruer1985}%
  \BibitemOpen
  \bibfield  {author} {\bibinfo {author} {\bibfnamefont {W.}~\bibnamefont
  {Kruer}}\ and\ \bibinfo {author} {\bibfnamefont {K.}~\bibnamefont
  {Estabrook}},\ }\href {\doibase 10.1063/1.865171} {\bibfield  {journal}
  {\bibinfo  {journal} {Physics of Fluids}\ }\textbf {\bibinfo {volume} {28}},\
  \bibinfo {pages} {430} (\bibinfo {year} {1985})}\BibitemShut {NoStop}%
\bibitem [{\citenamefont {Sheng}\ \emph
  {et~al.}(2002{\natexlab{b}})\citenamefont {Sheng}, \citenamefont {Mima},
  \citenamefont {Sentoku}, \citenamefont {Jovanovi\ifmmode~\acute{c}\else
  \'{c}\fi{}}, \citenamefont {Taguchi}, \citenamefont {Zhang},\ and\
  \citenamefont {Meyer-ter Vehn}}]{sheng2002-2}%
  \BibitemOpen
  \bibfield  {author} {\bibinfo {author} {\bibfnamefont {Z.-M.}\ \bibnamefont
  {Sheng}}, \bibinfo {author} {\bibfnamefont {K.}~\bibnamefont {Mima}},
  \bibinfo {author} {\bibfnamefont {Y.}~\bibnamefont {Sentoku}}, \bibinfo
  {author} {\bibfnamefont {M.~S.}\ \bibnamefont
  {Jovanovi\ifmmode~\acute{c}\else \'{c}\fi{}}}, \bibinfo {author}
  {\bibfnamefont {T.}~\bibnamefont {Taguchi}}, \bibinfo {author} {\bibfnamefont
  {J.}~\bibnamefont {Zhang}}, \ and\ \bibinfo {author} {\bibfnamefont
  {J.}~\bibnamefont {Meyer-ter Vehn}},\ }\href {\doibase
  10.1103/PhysRevLett.88.055004} {\bibfield  {journal} {\bibinfo  {journal}
  {Phys. Rev. Lett.}\ }\textbf {\bibinfo {volume} {88}},\ \bibinfo {pages}
  {055004} (\bibinfo {year} {2002}{\natexlab{b}})}\BibitemShut {NoStop}%
\bibitem [{\citenamefont {Thevenet}\ \emph {et~al.}(2016)\citenamefont
  {Thevenet}, \citenamefont {Leblanc}, \citenamefont {Kahaly}, \citenamefont
  {Vincenti}, \citenamefont {Vernier}, \citenamefont {Quere},\ and\
  \citenamefont {Faure}}]{thevenet2016}%
  \BibitemOpen
  \bibfield  {author} {\bibinfo {author} {\bibfnamefont {M.}~\bibnamefont
  {Thevenet}}, \bibinfo {author} {\bibfnamefont {A.}~\bibnamefont {Leblanc}},
  \bibinfo {author} {\bibfnamefont {S.}~\bibnamefont {Kahaly}}, \bibinfo
  {author} {\bibfnamefont {H.}~\bibnamefont {Vincenti}}, \bibinfo {author}
  {\bibfnamefont {A.}~\bibnamefont {Vernier}}, \bibinfo {author} {\bibfnamefont
  {F.}~\bibnamefont {Quere}}, \ and\ \bibinfo {author} {\bibfnamefont
  {J.}~\bibnamefont {Faure}},\ }\href {\doibase 10.1038/nphys3597} {\bibfield
  {journal} {\bibinfo  {journal} {Nature Physics}\ }\textbf {\bibinfo {volume}
  {12}},\ \bibinfo {pages} {355} (\bibinfo {year} {2016})}\BibitemShut
  {NoStop}%
\bibitem [{\citenamefont {Mao}\ \emph {et~al.}(2015)\citenamefont {Mao},
  \citenamefont {Chen}, \citenamefont {Huang}, \citenamefont {Ma},
  \citenamefont {Zhao}, \citenamefont {Li}, \citenamefont {Yan}, \citenamefont
  {Ma}, \citenamefont {Aeschlimann}, \citenamefont {Wei},\ and\ \citenamefont
  {Zhang}}]{mao2015}%
  \BibitemOpen
  \bibfield  {author} {\bibinfo {author} {\bibfnamefont {J.~Y.}\ \bibnamefont
  {Mao}}, \bibinfo {author} {\bibfnamefont {L.~M.}\ \bibnamefont {Chen}},
  \bibinfo {author} {\bibfnamefont {K.}~\bibnamefont {Huang}}, \bibinfo
  {author} {\bibfnamefont {Y.}~\bibnamefont {Ma}}, \bibinfo {author}
  {\bibfnamefont {J.~R.}\ \bibnamefont {Zhao}}, \bibinfo {author}
  {\bibfnamefont {D.~Z.}\ \bibnamefont {Li}}, \bibinfo {author} {\bibfnamefont
  {W.~C.}\ \bibnamefont {Yan}}, \bibinfo {author} {\bibfnamefont {J.~L.}\
  \bibnamefont {Ma}}, \bibinfo {author} {\bibfnamefont {M.}~\bibnamefont
  {Aeschlimann}}, \bibinfo {author} {\bibfnamefont {Z.~Y.}\ \bibnamefont
  {Wei}}, \ and\ \bibinfo {author} {\bibfnamefont {J.}~\bibnamefont {Zhang}},\
  }\href {\doibase 10.1063/1.4916636} {\bibfield  {journal} {\bibinfo
  {journal} {Applied Physics Letters}\ }\textbf {\bibinfo {volume} {106}},\
  \bibinfo {pages} {131105} (\bibinfo {year} {2015})}\BibitemShut {NoStop}%
\bibitem [{\citenamefont {Pukhov}\ \emph {et~al.}(1999)\citenamefont {Pukhov},
  \citenamefont {Sheng},\ and\ \citenamefont {ter Vehn}}]{pukhov1999}%
  \BibitemOpen
  \bibfield  {author} {\bibinfo {author} {\bibfnamefont {A.}~\bibnamefont
  {Pukhov}}, \bibinfo {author} {\bibfnamefont {Z.-M.}\ \bibnamefont {Sheng}}, \
  and\ \bibinfo {author} {\bibfnamefont {J.~M.}\ \bibnamefont {ter Vehn}},\
  }\href {\doibase 10.1063/1.873242} {\bibfield  {journal} {\bibinfo  {journal}
  {Physics of Plasmas}\ }\textbf {\bibinfo {volume} {6}},\ \bibinfo {pages}
  {2847} (\bibinfo {year} {1999})}\BibitemShut {NoStop}%
\bibitem [{\citenamefont {Gahn}\ \emph {et~al.}(1999)\citenamefont {Gahn},
  \citenamefont {Tsakiris}, \citenamefont {Pukhov}, \citenamefont {Meyer-ter
  Vehn}, \citenamefont {Pretzler}, \citenamefont {Thirolf}, \citenamefont
  {Habs},\ and\ \citenamefont {Witte}}]{gahn1999}%
  \BibitemOpen
  \bibfield  {author} {\bibinfo {author} {\bibfnamefont {C.}~\bibnamefont
  {Gahn}}, \bibinfo {author} {\bibfnamefont {G.~D.}\ \bibnamefont {Tsakiris}},
  \bibinfo {author} {\bibfnamefont {A.}~\bibnamefont {Pukhov}}, \bibinfo
  {author} {\bibfnamefont {J.}~\bibnamefont {Meyer-ter Vehn}}, \bibinfo
  {author} {\bibfnamefont {G.}~\bibnamefont {Pretzler}}, \bibinfo {author}
  {\bibfnamefont {P.}~\bibnamefont {Thirolf}}, \bibinfo {author} {\bibfnamefont
  {D.}~\bibnamefont {Habs}}, \ and\ \bibinfo {author} {\bibfnamefont {K.~J.}\
  \bibnamefont {Witte}},\ }\href {\doibase 10.1103/PhysRevLett.83.4772}
  {\bibfield  {journal} {\bibinfo  {journal} {Phys. Rev. Lett.}\ }\textbf
  {\bibinfo {volume} {83}},\ \bibinfo {pages} {4772} (\bibinfo {year}
  {1999})}\BibitemShut {NoStop}%
\bibitem [{\citenamefont {Mangles}\ \emph {et~al.}(2005)\citenamefont
  {Mangles}, \citenamefont {Walton}, \citenamefont {Tzoufras}, \citenamefont
  {Najmudin}, \citenamefont {Clarke}, \citenamefont {Dangor}, \citenamefont
  {Evans}, \citenamefont {Fritzler}, \citenamefont {Gopal}, \citenamefont
  {Hernandez-Gomez}, \citenamefont {Mori}, \citenamefont {Rozmus},
  \citenamefont {Tatarakis}, \citenamefont {Thomas}, \citenamefont {Tsung},
  \citenamefont {Wei},\ and\ \citenamefont {Krushelnick}}]{mangles2005}%
  \BibitemOpen
  \bibfield  {author} {\bibinfo {author} {\bibfnamefont {S.~P.~D.}\
  \bibnamefont {Mangles}}, \bibinfo {author} {\bibfnamefont {B.~R.}\
  \bibnamefont {Walton}}, \bibinfo {author} {\bibfnamefont {M.}~\bibnamefont
  {Tzoufras}}, \bibinfo {author} {\bibfnamefont {Z.}~\bibnamefont {Najmudin}},
  \bibinfo {author} {\bibfnamefont {R.~J.}\ \bibnamefont {Clarke}}, \bibinfo
  {author} {\bibfnamefont {A.~E.}\ \bibnamefont {Dangor}}, \bibinfo {author}
  {\bibfnamefont {R.~G.}\ \bibnamefont {Evans}}, \bibinfo {author}
  {\bibfnamefont {S.}~\bibnamefont {Fritzler}}, \bibinfo {author}
  {\bibfnamefont {A.}~\bibnamefont {Gopal}}, \bibinfo {author} {\bibfnamefont
  {C.}~\bibnamefont {Hernandez-Gomez}}, \bibinfo {author} {\bibfnamefont
  {W.~B.}\ \bibnamefont {Mori}}, \bibinfo {author} {\bibfnamefont
  {W.}~\bibnamefont {Rozmus}}, \bibinfo {author} {\bibfnamefont
  {M.}~\bibnamefont {Tatarakis}}, \bibinfo {author} {\bibfnamefont {A.~G.~R.}\
  \bibnamefont {Thomas}}, \bibinfo {author} {\bibfnamefont {F.~S.}\
  \bibnamefont {Tsung}}, \bibinfo {author} {\bibfnamefont {M.~S.}\ \bibnamefont
  {Wei}}, \ and\ \bibinfo {author} {\bibfnamefont {K.}~\bibnamefont
  {Krushelnick}},\ }\href {\doibase 10.1103/PhysRevLett.94.245001} {\bibfield
  {journal} {\bibinfo  {journal} {Phys. Rev. Lett.}\ }\textbf {\bibinfo
  {volume} {94}},\ \bibinfo {pages} {245001} (\bibinfo {year}
  {2005})}\BibitemShut {NoStop}%
\bibitem [{\citenamefont {Kneip}\ \emph {et~al.}(2008)\citenamefont {Kneip},
  \citenamefont {Nagel}, \citenamefont {Bellei}, \citenamefont {Bourgeois},
  \citenamefont {Dangor}, \citenamefont {Gopal}, \citenamefont {Heathcote},
  \citenamefont {Mangles}, \citenamefont {Marqu\`es}, \citenamefont
  {Maksimchuk}, \citenamefont {Nilson}, \citenamefont {Phuoc}, \citenamefont
  {Reed}, \citenamefont {Tzoufras}, \citenamefont {Tsung}, \citenamefont
  {Willingale}, \citenamefont {Mori}, \citenamefont {Rousse}, \citenamefont
  {Krushelnick},\ and\ \citenamefont {Najmudin}}]{kneip2008}%
  \BibitemOpen
  \bibfield  {author} {\bibinfo {author} {\bibfnamefont {S.}~\bibnamefont
  {Kneip}}, \bibinfo {author} {\bibfnamefont {S.~R.}\ \bibnamefont {Nagel}},
  \bibinfo {author} {\bibfnamefont {C.}~\bibnamefont {Bellei}}, \bibinfo
  {author} {\bibfnamefont {N.}~\bibnamefont {Bourgeois}}, \bibinfo {author}
  {\bibfnamefont {A.~E.}\ \bibnamefont {Dangor}}, \bibinfo {author}
  {\bibfnamefont {A.}~\bibnamefont {Gopal}}, \bibinfo {author} {\bibfnamefont
  {R.}~\bibnamefont {Heathcote}}, \bibinfo {author} {\bibfnamefont {S.~P.~D.}\
  \bibnamefont {Mangles}}, \bibinfo {author} {\bibfnamefont {J.~R.}\
  \bibnamefont {Marqu\`es}}, \bibinfo {author} {\bibfnamefont {A.}~\bibnamefont
  {Maksimchuk}}, \bibinfo {author} {\bibfnamefont {P.~M.}\ \bibnamefont
  {Nilson}}, \bibinfo {author} {\bibfnamefont {K.~T.}\ \bibnamefont {Phuoc}},
  \bibinfo {author} {\bibfnamefont {S.}~\bibnamefont {Reed}}, \bibinfo {author}
  {\bibfnamefont {M.}~\bibnamefont {Tzoufras}}, \bibinfo {author}
  {\bibfnamefont {F.~S.}\ \bibnamefont {Tsung}}, \bibinfo {author}
  {\bibfnamefont {L.}~\bibnamefont {Willingale}}, \bibinfo {author}
  {\bibfnamefont {W.~B.}\ \bibnamefont {Mori}}, \bibinfo {author}
  {\bibfnamefont {A.}~\bibnamefont {Rousse}}, \bibinfo {author} {\bibfnamefont
  {K.}~\bibnamefont {Krushelnick}}, \ and\ \bibinfo {author} {\bibfnamefont
  {Z.}~\bibnamefont {Najmudin}},\ }\href {\doibase
  10.1103/PhysRevLett.100.105006} {\bibfield  {journal} {\bibinfo  {journal}
  {Phys. Rev. Lett.}\ }\textbf {\bibinfo {volume} {100}},\ \bibinfo {pages}
  {105006} (\bibinfo {year} {2008})}\BibitemShut {NoStop}%
\bibitem [{\citenamefont {Li}\ \emph {et~al.}(2011)\citenamefont {Li},
  \citenamefont {Gu}, \citenamefont {Zhu}, \citenamefont {Li}, \citenamefont
  {Ban}, \citenamefont {Kong},\ and\ \citenamefont {Kawata}}]{liyy2011}%
  \BibitemOpen
  \bibfield  {author} {\bibinfo {author} {\bibfnamefont {Y.~Y.}\ \bibnamefont
  {Li}}, \bibinfo {author} {\bibfnamefont {Y.~J.}\ \bibnamefont {Gu}}, \bibinfo
  {author} {\bibfnamefont {Z.}~\bibnamefont {Zhu}}, \bibinfo {author}
  {\bibfnamefont {X.~F.}\ \bibnamefont {Li}}, \bibinfo {author} {\bibfnamefont
  {H.~Y.}\ \bibnamefont {Ban}}, \bibinfo {author} {\bibfnamefont
  {Q.}~\bibnamefont {Kong}}, \ and\ \bibinfo {author} {\bibfnamefont
  {S.}~\bibnamefont {Kawata}},\ }\href {\doibase 10.1063/1.3581062} {\bibfield
  {journal} {\bibinfo  {journal} {Physics of Plasmas}\ }\textbf {\bibinfo
  {volume} {18}},\ \bibinfo {pages} {053104} (\bibinfo {year}
  {2011})}\BibitemShut {NoStop}%
\bibitem [{\citenamefont {Zhang}\ \emph {et~al.}(2015)\citenamefont {Zhang},
  \citenamefont {Khudik},\ and\ \citenamefont {Shvets}}]{zhang2015}%
  \BibitemOpen
  \bibfield  {author} {\bibinfo {author} {\bibfnamefont {X.}~\bibnamefont
  {Zhang}}, \bibinfo {author} {\bibfnamefont {V.~N.}\ \bibnamefont {Khudik}}, \
  and\ \bibinfo {author} {\bibfnamefont {G.}~\bibnamefont {Shvets}},\ }\href
  {\doibase 10.1103/PhysRevLett.114.184801} {\bibfield  {journal} {\bibinfo
  {journal} {Phys. Rev. Lett.}\ }\textbf {\bibinfo {volume} {114}},\ \bibinfo
  {pages} {184801} (\bibinfo {year} {2015})}\BibitemShut {NoStop}%
\bibitem [{\citenamefont {Huang}\ \emph {et~al.}(2016)\citenamefont {Huang},
  \citenamefont {Robinson}, \citenamefont {Zhou}, \citenamefont {Qiao},
  \citenamefont {Liu}, \citenamefont {Ruan}, \citenamefont {He},\ and\
  \citenamefont {Norreys}}]{huang2016}%
  \BibitemOpen
  \bibfield  {author} {\bibinfo {author} {\bibfnamefont {T.~W.}\ \bibnamefont
  {Huang}}, \bibinfo {author} {\bibfnamefont {A.~P.~L.}\ \bibnamefont
  {Robinson}}, \bibinfo {author} {\bibfnamefont {C.~T.}\ \bibnamefont {Zhou}},
  \bibinfo {author} {\bibfnamefont {B.}~\bibnamefont {Qiao}}, \bibinfo {author}
  {\bibfnamefont {B.}~\bibnamefont {Liu}}, \bibinfo {author} {\bibfnamefont
  {S.~C.}\ \bibnamefont {Ruan}}, \bibinfo {author} {\bibfnamefont {X.~T.}\
  \bibnamefont {He}}, \ and\ \bibinfo {author} {\bibfnamefont {P.~A.}\
  \bibnamefont {Norreys}},\ }\href {\doibase 10.1103/PhysRevE.93.063203}
  {\bibfield  {journal} {\bibinfo  {journal} {Phys. Rev. E}\ }\textbf {\bibinfo
  {volume} {93}},\ \bibinfo {pages} {063203} (\bibinfo {year}
  {2016})}\BibitemShut {NoStop}%
\bibitem [{\citenamefont {Willingale}\ \emph {et~al.}(2011)\citenamefont
  {Willingale}, \citenamefont {Nilson}, \citenamefont {Thomas}, \citenamefont
  {Cobble}, \citenamefont {Craxton}, \citenamefont {Maksimchuk}, \citenamefont
  {Norreys}, \citenamefont {Sangster}, \citenamefont {Scott}, \citenamefont
  {Stoeckl}, \citenamefont {Zulick},\ and\ \citenamefont
  {Krushelnick}}]{louise2011}%
  \BibitemOpen
  \bibfield  {author} {\bibinfo {author} {\bibfnamefont {L.}~\bibnamefont
  {Willingale}}, \bibinfo {author} {\bibfnamefont {P.~M.}\ \bibnamefont
  {Nilson}}, \bibinfo {author} {\bibfnamefont {A.~G.~R.}\ \bibnamefont
  {Thomas}}, \bibinfo {author} {\bibfnamefont {J.}~\bibnamefont {Cobble}},
  \bibinfo {author} {\bibfnamefont {R.~S.}\ \bibnamefont {Craxton}}, \bibinfo
  {author} {\bibfnamefont {A.}~\bibnamefont {Maksimchuk}}, \bibinfo {author}
  {\bibfnamefont {P.~A.}\ \bibnamefont {Norreys}}, \bibinfo {author}
  {\bibfnamefont {T.~C.}\ \bibnamefont {Sangster}}, \bibinfo {author}
  {\bibfnamefont {R.~H.~H.}\ \bibnamefont {Scott}}, \bibinfo {author}
  {\bibfnamefont {C.}~\bibnamefont {Stoeckl}}, \bibinfo {author} {\bibfnamefont
  {C.}~\bibnamefont {Zulick}}, \ and\ \bibinfo {author} {\bibfnamefont
  {K.}~\bibnamefont {Krushelnick}},\ }\href {\doibase
  10.1103/PhysRevLett.106.105002} {\bibfield  {journal} {\bibinfo  {journal}
  {Phys. Rev. Lett.}\ }\textbf {\bibinfo {volume} {106}},\ \bibinfo {pages}
  {105002} (\bibinfo {year} {2011})}\BibitemShut {NoStop}%
\bibitem [{\citenamefont {Willingale}\ \emph {et~al.}(2013)\citenamefont
  {Willingale}, \citenamefont {Thomas}, \citenamefont {Nilson}, \citenamefont
  {Chen}, \citenamefont {Cobble}, \citenamefont {Craxton}, \citenamefont
  {Maksimchuk}, \citenamefont {Norreys}, \citenamefont {Sangster},
  \citenamefont {Scott}, \citenamefont {Stoeckl}, \citenamefont {Zulick},\ and\
  \citenamefont {Krushelnick}}]{louise2013}%
  \BibitemOpen
  \bibfield  {author} {\bibinfo {author} {\bibfnamefont {L.}~\bibnamefont
  {Willingale}}, \bibinfo {author} {\bibfnamefont {A.~G.~R.}\ \bibnamefont
  {Thomas}}, \bibinfo {author} {\bibfnamefont {P.~M.}\ \bibnamefont {Nilson}},
  \bibinfo {author} {\bibfnamefont {H.}~\bibnamefont {Chen}}, \bibinfo {author}
  {\bibfnamefont {J.}~\bibnamefont {Cobble}}, \bibinfo {author} {\bibfnamefont
  {R.~S.}\ \bibnamefont {Craxton}}, \bibinfo {author} {\bibfnamefont
  {A.}~\bibnamefont {Maksimchuk}}, \bibinfo {author} {\bibfnamefont {P.~A.}\
  \bibnamefont {Norreys}}, \bibinfo {author} {\bibfnamefont {T.~C.}\
  \bibnamefont {Sangster}}, \bibinfo {author} {\bibfnamefont {R.~H.~H.}\
  \bibnamefont {Scott}}, \bibinfo {author} {\bibfnamefont {C.}~\bibnamefont
  {Stoeckl}}, \bibinfo {author} {\bibfnamefont {C.}~\bibnamefont {Zulick}}, \
  and\ \bibinfo {author} {\bibfnamefont {K.}~\bibnamefont {Krushelnick}},\
  }\href {http://stacks.iop.org/1367-2630/15/i=2/a=025023} {\bibfield
  {journal} {\bibinfo  {journal} {New Journal of Physics}\ }\textbf {\bibinfo
  {volume} {15}},\ \bibinfo {pages} {025023} (\bibinfo {year}
  {2013})}\BibitemShut {NoStop}%
\bibitem [{\citenamefont {Alfv\'en}(1939)}]{alfven1939}%
  \BibitemOpen
  \bibfield  {author} {\bibinfo {author} {\bibfnamefont {H.}~\bibnamefont
  {Alfv\'en}},\ }\href {\doibase 10.1103/PhysRev.55.425} {\bibfield  {journal}
  {\bibinfo  {journal} {Phys. Rev.}\ }\textbf {\bibinfo {volume} {55}},\
  \bibinfo {pages} {425} (\bibinfo {year} {1939})}\BibitemShut {NoStop}%
\bibitem [{\citenamefont {Dodin}\ and\ \citenamefont
  {Fisch}(2006)}]{dodin2006}%
  \BibitemOpen
  \bibfield  {author} {\bibinfo {author} {\bibfnamefont {I.~Y.}\ \bibnamefont
  {Dodin}}\ and\ \bibinfo {author} {\bibfnamefont {N.~J.}\ \bibnamefont
  {Fisch}},\ }\href {\doibase 10.1063/1.2358970} {\bibfield  {journal}
  {\bibinfo  {journal} {Physics of Plasmas}\ }\textbf {\bibinfo {volume}
  {13}},\ \bibinfo {pages} {103104} (\bibinfo {year} {2006})}\BibitemShut
  {NoStop}%
\bibitem [{\citenamefont {Sentoku}\ \emph {et~al.}(1999)\citenamefont
  {Sentoku}, \citenamefont {Ruhl}, \citenamefont {Mima}, \citenamefont
  {Kodama}, \citenamefont {Tanaka},\ and\ \citenamefont
  {Kishimoto}}]{sentoku1999}%
  \BibitemOpen
  \bibfield  {author} {\bibinfo {author} {\bibfnamefont {Y.}~\bibnamefont
  {Sentoku}}, \bibinfo {author} {\bibfnamefont {H.}~\bibnamefont {Ruhl}},
  \bibinfo {author} {\bibfnamefont {K.}~\bibnamefont {Mima}}, \bibinfo {author}
  {\bibfnamefont {R.}~\bibnamefont {Kodama}}, \bibinfo {author} {\bibfnamefont
  {K.}~\bibnamefont {Tanaka}}, \ and\ \bibinfo {author} {\bibfnamefont
  {Y.}~\bibnamefont {Kishimoto}},\ }\href {\doibase 10.1063/1.873243}
  {\bibfield  {journal} {\bibinfo  {journal} {Physics of Plasmas}\ }\textbf
  {\bibinfo {volume} {6}},\ \bibinfo {pages} {2855} (\bibinfo {year}
  {1999})}\BibitemShut {NoStop}%
\bibitem [{\citenamefont {Liu}\ \emph {et~al.}(2013)\citenamefont {Liu},
  \citenamefont {Wang}, \citenamefont {Liu}, \citenamefont {Fu}, \citenamefont
  {Xu}, \citenamefont {Yan},\ and\ \citenamefont {He}}]{liub2013}%
  \BibitemOpen
  \bibfield  {author} {\bibinfo {author} {\bibfnamefont {B.}~\bibnamefont
  {Liu}}, \bibinfo {author} {\bibfnamefont {H.~Y.}\ \bibnamefont {Wang}},
  \bibinfo {author} {\bibfnamefont {J.}~\bibnamefont {Liu}}, \bibinfo {author}
  {\bibfnamefont {L.~B.}\ \bibnamefont {Fu}}, \bibinfo {author} {\bibfnamefont
  {Y.~J.}\ \bibnamefont {Xu}}, \bibinfo {author} {\bibfnamefont {X.~Q.}\
  \bibnamefont {Yan}}, \ and\ \bibinfo {author} {\bibfnamefont {X.~T.}\
  \bibnamefont {He}},\ }\href {\doibase 10.1103/PhysRevLett.110.045002}
  {\bibfield  {journal} {\bibinfo  {journal} {Phys. Rev. Lett.}\ }\textbf
  {\bibinfo {volume} {110}},\ \bibinfo {pages} {045002} (\bibinfo {year}
  {2013})}\BibitemShut {NoStop}%
\bibitem [{\citenamefont {Tzoufras}\ \emph {et~al.}(2008)\citenamefont
  {Tzoufras}, \citenamefont {Lu}, \citenamefont {Tsung}, \citenamefont {Huang},
  \citenamefont {Mori}, \citenamefont {Katsouleas}, \citenamefont {Vieira},
  \citenamefont {Fonseca},\ and\ \citenamefont {Silva}}]{tzoufras2008}%
  \BibitemOpen
  \bibfield  {author} {\bibinfo {author} {\bibfnamefont {M.}~\bibnamefont
  {Tzoufras}}, \bibinfo {author} {\bibfnamefont {W.}~\bibnamefont {Lu}},
  \bibinfo {author} {\bibfnamefont {F.~S.}\ \bibnamefont {Tsung}}, \bibinfo
  {author} {\bibfnamefont {C.}~\bibnamefont {Huang}}, \bibinfo {author}
  {\bibfnamefont {W.~B.}\ \bibnamefont {Mori}}, \bibinfo {author}
  {\bibfnamefont {T.}~\bibnamefont {Katsouleas}}, \bibinfo {author}
  {\bibfnamefont {J.}~\bibnamefont {Vieira}}, \bibinfo {author} {\bibfnamefont
  {R.~A.}\ \bibnamefont {Fonseca}}, \ and\ \bibinfo {author} {\bibfnamefont
  {L.~O.}\ \bibnamefont {Silva}},\ }\href {\doibase
  10.1103/PhysRevLett.101.145002} {\bibfield  {journal} {\bibinfo  {journal}
  {Phys. Rev. Lett.}\ }\textbf {\bibinfo {volume} {101}},\ \bibinfo {pages}
  {145002} (\bibinfo {year} {2008})}\BibitemShut {NoStop}%
\bibitem [{\citenamefont {Bostedt}\ \emph {et~al.}(2016)\citenamefont
  {Bostedt}, \citenamefont {Boutet}, \citenamefont {Fritz}, \citenamefont
  {Huang}, \citenamefont {Lee}, \citenamefont {Lemke}, \citenamefont {Robert},
  \citenamefont {Schlotter}, \citenamefont {Turner},\ and\ \citenamefont
  {Williams}}]{LCLS}%
  \BibitemOpen
  \bibfield  {author} {\bibinfo {author} {\bibfnamefont {C.}~\bibnamefont
  {Bostedt}}, \bibinfo {author} {\bibfnamefont {S.}~\bibnamefont {Boutet}},
  \bibinfo {author} {\bibfnamefont {D.~M.}\ \bibnamefont {Fritz}}, \bibinfo
  {author} {\bibfnamefont {Z.}~\bibnamefont {Huang}}, \bibinfo {author}
  {\bibfnamefont {H.~J.}\ \bibnamefont {Lee}}, \bibinfo {author} {\bibfnamefont
  {H.~T.}\ \bibnamefont {Lemke}}, \bibinfo {author} {\bibfnamefont
  {A.}~\bibnamefont {Robert}}, \bibinfo {author} {\bibfnamefont {W.~F.}\
  \bibnamefont {Schlotter}}, \bibinfo {author} {\bibfnamefont {J.~J.}\
  \bibnamefont {Turner}}, \ and\ \bibinfo {author} {\bibfnamefont {G.~J.}\
  \bibnamefont {Williams}},\ }\href {\doibase 10.1103/RevModPhys.88.015007}
  {\bibfield  {journal} {\bibinfo  {journal} {Rev. Mod. Phys.}\ }\textbf
  {\bibinfo {volume} {88}},\ \bibinfo {pages} {015007} (\bibinfo {year}
  {2016})}\BibitemShut {NoStop}%
\bibitem [{\citenamefont {Tabak}\ \emph {et~al.}(1994)\citenamefont {Tabak},
  \citenamefont {Hammer}, \citenamefont {Glinsky}, \citenamefont {Kruer},
  \citenamefont {Wilks}, \citenamefont {Woodworth}, \citenamefont {Campbell},
  \citenamefont {Perry},\ and\ \citenamefont {Mason}}]{tabak1994}%
  \BibitemOpen
  \bibfield  {author} {\bibinfo {author} {\bibfnamefont {M.}~\bibnamefont
  {Tabak}}, \bibinfo {author} {\bibfnamefont {J.}~\bibnamefont {Hammer}},
  \bibinfo {author} {\bibfnamefont {M.}~\bibnamefont {Glinsky}}, \bibinfo
  {author} {\bibfnamefont {W.}~\bibnamefont {Kruer}}, \bibinfo {author}
  {\bibfnamefont {S.}~\bibnamefont {Wilks}}, \bibinfo {author} {\bibfnamefont
  {J.}~\bibnamefont {Woodworth}}, \bibinfo {author} {\bibfnamefont
  {E.}~\bibnamefont {Campbell}}, \bibinfo {author} {\bibfnamefont
  {M.}~\bibnamefont {Perry}}, \ and\ \bibinfo {author} {\bibfnamefont
  {R.}~\bibnamefont {Mason}},\ }\href {\doibase 10.1063/1.870664} {\bibfield
  {journal} {\bibinfo  {journal} {Physics of Plasmas}\ }\textbf {\bibinfo
  {volume} {1}},\ \bibinfo {pages} {1626} (\bibinfo {year} {1994})},\ \bibinfo
  {note} {35th Annual Meeting of the Division-of-Plasma-Physics of the
  American-Physical-Society, St Louis, MO, Nov 01-05, 1993}\BibitemShut
  {NoStop}%
\bibitem [{\citenamefont {Tanaka}\ \emph {et~al.}(2005)\citenamefont {Tanaka},
  \citenamefont {Yabuuchi}, \citenamefont {Sato}, \citenamefont {Kodama},
  \citenamefont {Kitagawa}, \citenamefont {Takahashi}, \citenamefont {Ikeda},
  \citenamefont {Honda},\ and\ \citenamefont {Okuda}}]{tanaka2005}%
  \BibitemOpen
  \bibfield  {author} {\bibinfo {author} {\bibfnamefont {K.~A.}\ \bibnamefont
  {Tanaka}}, \bibinfo {author} {\bibfnamefont {T.}~\bibnamefont {Yabuuchi}},
  \bibinfo {author} {\bibfnamefont {T.}~\bibnamefont {Sato}}, \bibinfo {author}
  {\bibfnamefont {R.}~\bibnamefont {Kodama}}, \bibinfo {author} {\bibfnamefont
  {Y.}~\bibnamefont {Kitagawa}}, \bibinfo {author} {\bibfnamefont
  {T.}~\bibnamefont {Takahashi}}, \bibinfo {author} {\bibfnamefont
  {T.}~\bibnamefont {Ikeda}}, \bibinfo {author} {\bibfnamefont
  {Y.}~\bibnamefont {Honda}}, \ and\ \bibinfo {author} {\bibfnamefont
  {S.}~\bibnamefont {Okuda}},\ }\href {\doibase 10.1063/1.1824371} {\bibfield
  {journal} {\bibinfo  {journal} {Review of Scientific Instruments}\ }\textbf
  {\bibinfo {volume} {76}},\ \bibinfo {pages} {013507} (\bibinfo {year}
  {2005})}\BibitemShut {NoStop}%
\bibitem [{\citenamefont {Briesmeister}(2000)}]{Briesmeister2000}%
  \BibitemOpen
  \bibfield  {author} {\bibinfo {author} {\bibfnamefont {J.~F.}\ \bibnamefont
  {Briesmeister}},\ }\href@noop {} {\emph {\bibinfo {title} {{MCNP: A General
  Monte Carlo N-Particle Transport Code}}}}\ (\bibinfo  {publisher} {Los Alamos
  National Laboratory Report No. LA-13709-M},\ \bibinfo {year}
  {2000})\BibitemShut {NoStop}%
\bibitem [{\citenamefont {Maddox}\ \emph {et~al.}(2011)\citenamefont {Maddox},
  \citenamefont {Park}, \citenamefont {Remington}, \citenamefont {Izumi},
  \citenamefont {Chen}, \citenamefont {Chen}, \citenamefont {Kimminau},
  \citenamefont {Ali}, \citenamefont {Haugh},\ and\ \citenamefont
  {Ma}}]{maddox2011}%
  \BibitemOpen
  \bibfield  {author} {\bibinfo {author} {\bibfnamefont {B.~R.}\ \bibnamefont
  {Maddox}}, \bibinfo {author} {\bibfnamefont {H.~S.}\ \bibnamefont {Park}},
  \bibinfo {author} {\bibfnamefont {B.~A.}\ \bibnamefont {Remington}}, \bibinfo
  {author} {\bibfnamefont {N.}~\bibnamefont {Izumi}}, \bibinfo {author}
  {\bibfnamefont {S.}~\bibnamefont {Chen}}, \bibinfo {author} {\bibfnamefont
  {C.}~\bibnamefont {Chen}}, \bibinfo {author} {\bibfnamefont {G.}~\bibnamefont
  {Kimminau}}, \bibinfo {author} {\bibfnamefont {Z.}~\bibnamefont {Ali}},
  \bibinfo {author} {\bibfnamefont {M.~J.}\ \bibnamefont {Haugh}}, \ and\
  \bibinfo {author} {\bibfnamefont {Q.}~\bibnamefont {Ma}},\ }\href {\doibase
  10.1063/1.3531979} {\bibfield  {journal} {\bibinfo  {journal} {Review of
  Scientific Instruments}\ }\textbf {\bibinfo {volume} {82}},\ \bibinfo {pages}
  {023111} (\bibinfo {year} {2011})}\BibitemShut {NoStop}%
\bibitem [{\citenamefont {Arber}\ \emph {et~al.}(2015)\citenamefont {Arber},
  \citenamefont {Bennett}, \citenamefont {Brady}, \citenamefont
  {Lawrence-Douglas}, \citenamefont {Ramsay}, \citenamefont {Sircombe},
  \citenamefont {Gillies}, \citenamefont {Evans}, \citenamefont {Schmitz},
  \citenamefont {Bell},\ and\ \citenamefont {Ridgers}}]{arber2015}%
  \BibitemOpen
  \bibfield  {author} {\bibinfo {author} {\bibfnamefont {T.~D.}\ \bibnamefont
  {Arber}}, \bibinfo {author} {\bibfnamefont {K.}~\bibnamefont {Bennett}},
  \bibinfo {author} {\bibfnamefont {C.~S.}\ \bibnamefont {Brady}}, \bibinfo
  {author} {\bibfnamefont {A.}~\bibnamefont {Lawrence-Douglas}}, \bibinfo
  {author} {\bibfnamefont {M.~G.}\ \bibnamefont {Ramsay}}, \bibinfo {author}
  {\bibfnamefont {N.~J.}\ \bibnamefont {Sircombe}}, \bibinfo {author}
  {\bibfnamefont {P.}~\bibnamefont {Gillies}}, \bibinfo {author} {\bibfnamefont
  {R.~G.}\ \bibnamefont {Evans}}, \bibinfo {author} {\bibfnamefont
  {H.}~\bibnamefont {Schmitz}}, \bibinfo {author} {\bibfnamefont {A.~R.}\
  \bibnamefont {Bell}}, \ and\ \bibinfo {author} {\bibfnamefont {C.~P.}\
  \bibnamefont {Ridgers}},\ }\href
  {http://stacks.iop.org/0741-3335/57/i=11/a=113001} {\bibfield  {journal}
  {\bibinfo  {journal} {Plasma Physics and Controlled Fusion}\ }\textbf
  {\bibinfo {volume} {57}},\ \bibinfo {pages} {1} (\bibinfo {year}
  {2015})}\BibitemShut {NoStop}%
\bibitem [{\citenamefont {Ramis}\ \emph {et~al.}(2009)\citenamefont {Ramis},
  \citenamefont {ter Vehn},\ and\ \citenamefont {Ramírez}}]{Ramis2009977}%
  \BibitemOpen
  \bibfield  {author} {\bibinfo {author} {\bibfnamefont {R.}~\bibnamefont
  {Ramis}}, \bibinfo {author} {\bibfnamefont {J.~M.}\ \bibnamefont {ter Vehn}},
  \ and\ \bibinfo {author} {\bibfnamefont {J.}~\bibnamefont {Ramírez}},\
  }\href {\doibase http://dx.doi.org/10.1016/j.cpc.2008.12.033} {\bibfield
  {journal} {\bibinfo  {journal} {Computer Physics Communications}\ }\textbf
  {\bibinfo {volume} {180}},\ \bibinfo {pages} {977 } (\bibinfo {year}
  {2009})}\BibitemShut {NoStop}%
\bibitem [{\citenamefont {Shaw}\ \emph {et~al.}(2017)\citenamefont {Shaw},
  \citenamefont {Lemos}, \citenamefont {Amorim}, \citenamefont
  {Vafaei-Najafabadi}, \citenamefont {Marsh}, \citenamefont {Tsung},
  \citenamefont {Mori},\ and\ \citenamefont {Joshi}}]{shaw2017}%
  \BibitemOpen
  \bibfield  {author} {\bibinfo {author} {\bibfnamefont {J.~L.}\ \bibnamefont
  {Shaw}}, \bibinfo {author} {\bibfnamefont {N.}~\bibnamefont {Lemos}},
  \bibinfo {author} {\bibfnamefont {L.~D.}\ \bibnamefont {Amorim}}, \bibinfo
  {author} {\bibfnamefont {N.}~\bibnamefont {Vafaei-Najafabadi}}, \bibinfo
  {author} {\bibfnamefont {K.~A.}\ \bibnamefont {Marsh}}, \bibinfo {author}
  {\bibfnamefont {F.~S.}\ \bibnamefont {Tsung}}, \bibinfo {author}
  {\bibfnamefont {W.~B.}\ \bibnamefont {Mori}}, \ and\ \bibinfo {author}
  {\bibfnamefont {C.}~\bibnamefont {Joshi}},\ }\href {\doibase
  10.1103/PhysRevLett.118.064801} {\bibfield  {journal} {\bibinfo  {journal}
  {Phys. Rev. Lett.}\ }\textbf {\bibinfo {volume} {118}},\ \bibinfo {pages}
  {064801} (\bibinfo {year} {2017})}\BibitemShut {NoStop}%
\end{thebibliography}%

\end{document}